\documentclass[superscriptaddress,11pt,nofootinbib,standalone]{revtex4-1}

\usepackage{bbm,bm,mathtools,slashed}
\usepackage{graphicx,color}
\usepackage{hyperref}
\usepackage[T1]{fontenc}
\usepackage{enumerate}
\usepackage[all]{xy}
\usepackage{amsmath,amssymb,amsthm}
\usepackage{ascmac}
\usepackage{braket}
\usepackage{extarrows}
\usepackage{cases}
\usepackage{tikz}
\usetikzlibrary{intersections, calc}

\newcommand{\calM}{\mathcal{M}}

\newcommand{\barpsi}{\bar{\psi}}
\newcommand{\Dirac}[1]{D_{#1}}
\newcommand{\calD}{\mathcal{D}}

\newcommand{\gam}[1]{\gamma_{#1}}
\newcommand{\ide}{\bm{1}_{N}}
\newcommand{\idef}{\bm{1}_{4}}

\newcommand{\ot}{\otimes}
\newcommand{\bigot}{\bigotimes}
\newcommand{\Pmat}{P_{N}}

\newcommand{\hmu}{\hat{\mu}}
\newcommand{\bpmat}{\begin{pmatrix}}
\newcommand{\epmat}{\end{pmatrix}}
\newcommand{\U}[1]{U_{n,#1}}
\newcommand{\diag}[1]{\mathrm{Diag} \left[ #1 \right]}

\newcommand{\sumk}{\sum_{k_{4},k_{3},k_{2},k_{1}}}
\newcommand{\ketk}{\ket{k_{4},k_{3},k_{2},k_{1}}}
\newcommand{\brak}{\bra{k_{4},k_{3},k_{2},k_{1}}}
\newcommand{\proj}{\mathcal{P}}
\newcommand{\Uni}{\mathcal{U}}

\newcommand{\calV}{\mathcal{V}}
\newcommand{\init}{\mathrm{init}}
\newcommand{\ter}{\mathrm{ter}}

\newtheorem{definition}{Definition.}

\begin{document}

\title{
Lattice fermions as spectral graphs
}

\author{Jun Yumoto}
\email{d8521007(at)s.akita-u.ac.jp}
\address{Department of Mathematical Science, Akita University, Akita 010-8502, Japan}

\author{Tatsuhiro Misumi}
\email{misumi(at)phys.kindai.ac.jp}
\address{Department of Physics, Kindai University,  Osaka 577-8502, Japan}
\address{Research and Education Center for Natural Sciences, Keio University, Kanagawa 223-8521, Japan}

\begin{abstract}
We study lattice fermions from the viewpoint of spectral graph theory (SGT).
We find that a fermion defined on a certain lattice is identified as a spectral graph.
SGT helps us investigate the number of zero eigenvalues of lattice Dirac operators even on the non-torus and non-regular lattice, leading to understanding of the number of fermion species (doublers) on lattices with arbitrary topologies.  
The procedure of application of SGT to lattice fermions is summarized as follows:
(1) One investigates a spectral graph corresponding to a lattice fermion.
(2) One obtains a matrix corresponding to the graph.
(3) One finds zero eigenvalues of the matrix by use of the discrete Fourier transformation (DFT).
(4) By taking an infinite-volume and continuum limits, one finds the number of species.
We apply this procedure to the known lattice fermion formulations including Naive fermions, Wilson fermions and Domain-wall fermions, and reproduce the known fact on the number of species.
We also apply it to the lattice fermion on the discretized four-dimensional hyperball and discuss the number of fermion species on the bulk.
In the end of the paper, we discuss the application of the analysis to lattice fermions on generic lattices with arbitrary topologies, which could lead to constructing a new theorem regarding the number of species. 
\end{abstract}

\maketitle

\newpage

\tableofcontents

\newpage


\section{Introduction}
\label{sec:Intro}

Non-perturbative physics of quantum field theories has been investigated by lattice gauge theory \cite{Wilson:1974sk, Creutz:1980zw}.  
Despite of the accomplishments obtained by the lattice Monte-Carlo simulation, 
there are still troublesome problems in lattice fermion formulations \cite{Karsten:1980wd, Nielsen:1980rz, Nielsen:1981xu, Nielsen:1981hk}, including the reconcilement of chiral symmetry and numerical efficiency, the realization of a single Weyl fermion, and the sign problem of the quark determinant. The known lattice fermions, including Wilson, staggered and domain-wall (overlap) fermions, have their own shortcomings \cite{Wilson:1975id, Kaplan:1992bt, Shamir:1993zy, Furman:1994ky, Neuberger:1998wv,Ginsparg:1981bj, Kogut:1974ag, Susskind:1976jm,
Kawamoto:1981hw,Sharatchandra:1981si,Golterman:1984cy,Golterman:1985dz,Kilcup:1986dg}.
Other lattice fermion formulations have been investigated, but they also have their own drawbacks: The generalized Wilson fermions based on flavored-mass terms \cite{Bietenholz:1999km,Creutz:2010bm,Durr:2010ch,Durr:2012dw,
Misumi:2012eh,Cho:2013yha,Cho:2015ffa,Durr:2017wfi} and the staggered-Wilson fermions \cite{Golterman:1984cy, Adams:2009eb,Adams:2010gx,Hoelbling:2010jw, deForcrand:2011ak,Creutz:2011cd,Misumi:2011su,Follana:2011kh,deForcrand:2012bm,Misumi:2012sp,Misumi:2012eh,Durr:2013gp,Hoelbling:2016qfv,Zielinski:2017pko} can reduce numerical costs and discretizartion errors. However, their implementation to lattice QCD is complicated and eventually leads to costs.
The minimally doubled fermion \cite{Karsten:1981gd,Wilczek:1987kw,Creutz:2007af,Borici:2007kz,Bedaque:2008xs,Bedaque:2008jm,
Capitani:2009yn,Kimura:2009qe,Kimura:2009di,Creutz:2010cz,Capitani:2010nn,Tiburzi:2010bm,Kamata:2011jn,Misumi:2012uu,Misumi:2012ky,Capitani:2013zta,Capitani:2013iha,Misumi:2013maa,Weber:2013tfa,Weber:2017eds,Durr:2020yqa} with $U(1)$ chiral symmetry yields only two species. The shortcoming of this setup is the explicit breaking of part of the lattice discrete symmetry.
The central-branch Wilson fermion \cite{Kimura:2011ik, Creutz:2011cd,Misumi:2012eh,Chowdhury:2013ux, Misumi:2020eyx} has the extra $U(1)_{\overline V}$ symmetry prohibiting additive mass renormalization, but its two-flavor version loses part of hypercubic symmetry.

One of new avenues in the study of lattice fermions is to study fermions on non-torus lattices with different topologies. However, we are so far unfamiliar with the number of species on such lattices since the Nielsen-Ninomiya's no-go theorem can be only applied to the lattice fermions defined on the torus\footnote{Some may state that the Nielsen-Ninomiya's no-go theorem can be applied to fermions on non-torus lattices. However, this no-go theorem cannot explain the empirical fact that the number of fermion species is two on the discretized two-sphere.}. Toward understanding the number of species of fermions on lattices with arbitrary topologies, we need to develop the way how to figure out the number of zero eigenvalues of lattice Dirac operators without using the momentum representation since the non-torus lattices do not have discrete translational symmetry.

In this paper, we discuss lattice fermions in terms of spectral graph theory (SGT) \cite{west2001introduction,bondy1976graph,mieghem_2010,Watts1998} and obtain a new insight.
The relation between supersymmetric lattice gauge theories (matrix models) and graph theories has been recently discussed in \cite{Ohta:2020ygi,Ohta:2021xty}, and the relation among Dirac-Kaehler fermions, Euler characteristics of background spaces and quantum anomalies has been discussed in \cite{Catterall:2018lkj,Butt:2021brl}.
We now focus on the spectral graph theory (SGT) and the discrete Fourier transformation (DFT), which tell us the relation among a graph, the associated matrix and its eigenvalues. 
We find that a Dirac operator of fermions defined on a certain lattice is identified as a matrix of the spectral graph corresponding to the lattice fermion.
SGT and DFT clarifies the number of zero eigenvalues of lattice Dirac operators, leading to understanding of the number of doublers on lattices with arbitrary topologies.  
Our arguments are summarized as follows;
First of all, a lattice fermion on a certain lattice corresponds to one spectral graph.
Secondly, the Dirac operator of the lattice fermion can be treated as a matrix corresponding to the graph. 
Thus, once we have a certain lattice and a Dirac operator defined on the lattice,
we can derive the corresponding matrix and find the number of zero eigenvalues by use of the discrete Fourier transformation (DFT).
Finally, by taking an infinite-volume and continuum limits, we can study the number of doublers.
We apply this procedure to the known fermion formulations including Naive fermion, Wilson fermion and Domain-wall fermion defined on the regular lattices, although the advantage of our technique is its use for non-torus lattices.
We then apply it to the lattice fermion on the discretized four-dimensional hyperball and study the number of doublers on the bulk. We also discuss its application to hypersphere and discuss the possibility of a new theorem regarding the number of doublers on the lattice with arbitrary topologies. 

The main purpose of this work is to introduce the importance of spectral graph theory to the lattice field theory community. Although most of the results obtained in this paper are reproductions of the known results, the method we use here is new and it could be applied to the nontrivial cases in future.

This paper is constructed as follows:
In Sec.~\ref{sec:SGT} we introduce the spectral graph theory.
In Sec.~\ref{sec:naive} we discuss the naive fermion in terms of SGT and reproduce the known results.
In Sec.~\ref{sec:Wilson} and Sec.~\ref{sec:DW} we discuss Wilson and domain-wall fermions in terms of SGT and reproduce the known results.
In Sec.~\ref{sec:lf_on_Bd}, as the first non-trivial application, we apply SGT analysis to the lattice fermions on four-dimensional hyperball.
Sec.~\ref{sec:SD} is devoted to the summary and discussion.

\section{Spectral graph theory}
\label{sec:SGT}


In the section, we introduce basic concepts of Spectral Graph Theory (SGT) for the purpose of discussing lattice fermions from a new viewpoint. 
We will also present some examples to show the importance of the concept. 
Let us start with the definition of a graph \cite{west2001introduction,bondy1976graph,mieghem_2010,Watts1998}.
\begin{definition}[graph]
	A graph $G$ is a pair $G = (V,E)$, where $V$ is a set of vertices of the graph and $E$ is a set of edges of the graph.
\end{definition}

\begin{figure}[htpb]
\centering
	\includegraphics[clip,
		height=4cm]{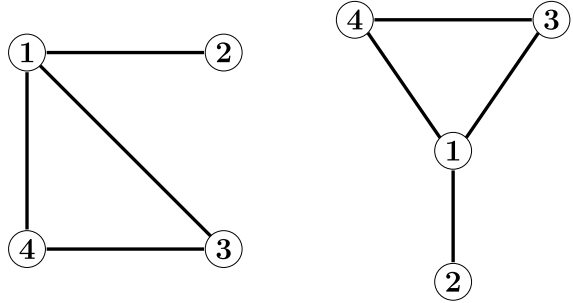}
\caption{These examples are graphs having a pair $G=(V,E)$ with $V = \left\{ 1,2,3,4 \right\}$ and $E = \left\{ \{ 1,2 \}, \{ 1,3 \}, \{ 1,4 \}, \{ 3,4 \} \right\}$.}
\label{graph:undirected}
\end{figure}

As examples, we exhibit two graphs in Fig.~\ref{graph:undirected} with $V = \left\{ 1,2,3,4 \right\}$ and $E = \left\{ \{ 1,2 \}, \{ 1,3 \}, \{ 1,4 \}, \{ 3,4 \} \right\}$. Here, $\{i,j\}$ stands for an edge from $i$ to $j$. 
Note that two graphs in Fig.~\ref{graph:undirected} have the common pair $G = (V,E)$ and they are the identical graphs although their appearances are different.
They have no directed edges, which will be discussed soon, thus any two vertex in any element of $E$ in the examples can be commutated, so we find $\{ i, j \} = \{ j, i \}$.

\begin{definition}[directed graph or digraph]
	A directed graph (or digraph) is a pair $(V,E)$ of sets of vertices and edges together with two maps $\init : E \to V$ and $\ter : E \to V$. The two maps are assigned to every edge $e_{ij}$ with an initial vertex $\init (e_{ij}) = v_{i} \in V$ and a terminal vertex $\ter(e_{ij}) = v_{j} \in V$. The edge $e_{ij}$ is said to be directed from $\init (e_{ij})$ to $\ter (e_{ij})$. If $\init(e_{ij}) = \ter(e_{ij})$, the edge $e_{ij}$ is called a loop.
\end{definition}

\begin{figure}[htpb]
\centering
	\includegraphics[
	clip,height=4cm]{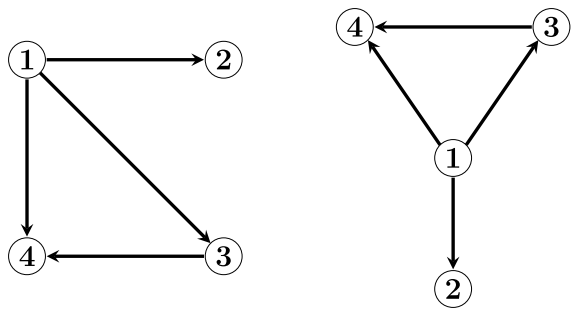}
\caption{These examples are graphs having a pair $(V,E)$ with $V = \left\{ 1,2,3,4 \right\}$ and $E = \left\{ \{ 1,2 \}, \{ 1,3 \}, \{ 1,4 \}, \{ 3,4 \} \right\}$. The initial vertices of edges are $\init(\{1,2\}) = 1, \init(\{1,3\}) = 1, \init(\{1,4\}) = 1, \init(\{3,4\}) = 3$, while the terminal vertices are $\init(\{1,2\}) = 2, \init(\{1,3\}) = 3, \init(\{1,4\}) = 4, \init(\{3,4\}) = 4$.}
\label{graph:directed}
\end{figure}

As examples, two graphs in Fig.~\ref{graph:directed} are digraphs with $V = \left\{ 1,2,3,4 \right\}$ and $E = \left\{ \{ 1,2 \}, \{ 1,3 \}, \{ 1,4 \}, \{ 3,4 \} \right\}$. Then initial vertex and terminal vertex are assigned as $\init(\{i,j\}) = i, \ter(\{i,j\}) = j$.

\begin{figure}[htpb]
\centering
	\includegraphics[
		clip,height=4cm]{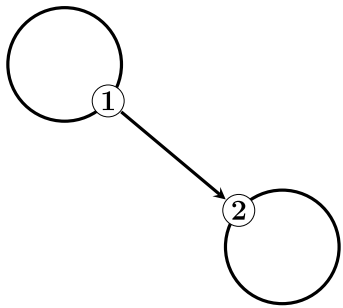}
\caption{A digraph with two loops.}
\label{graph:loop}
\end{figure}

As another example we show Fig.~\ref{graph:loop}, which is a digraph with loop. Note that vertices in any loop can be commutated since they are the identical vertex. In general, two vertices in a directed edge can not be commutated $\{ i,j \} \neq \{ j,i \}$.

\begin{definition}[weighted graph]
	The weighted graph has a value (the weight) for each edge in a graph or a digraph.
\end{definition}

\begin{figure}[htpb]
\centering
	\includegraphics[
		clip,height=4cm]{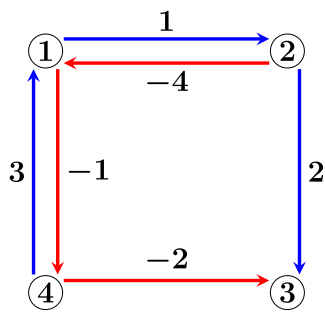}
\caption{This digraph is a weighted graph. Blue edges in the graph are those with positive weights, while red edges are those with negative weights.}
\label{graph:weighted}
\end{figure}

As an example, we show a digraph in Fig.~\ref{graph:weighted}. It is a weighted graph, each of whose edge has a weight.

\begin{definition}[adjacency matrix]
	The adjacency matrix $A$ of a graph is the $|V| \times |V|$ matrix given by
	\begin{equation}
		A_{ij} = \begin{cases}
			w_{ij}	& \mbox{if there is a edge from $i$ to $j$} \\
			0		& \mbox{otherwise}
		\end{cases}\,,
	\end{equation}
	where $w_{ij}$ is the weight of an edge from $i$ to $j$.
\end{definition}

As an example we exhibit an adjacency matrix $A$ of a graph in Fig.~\ref{graph:weighted}
\begin{equation}
	A = \begin{pmatrix}
		0 & 1 & 0 & -1 \\
		-4 & 0 & 2 & 0 \\
		0 & 0 & 0 & 0 \\
		3 & 0 & -2 & 0
	\end{pmatrix}\,.
\end{equation}
This matrix is asymmetric. In general, the adjacency matrix of a directed graph is asymmetric since the existence of an edge from $i$ to $j$ does not necessarily imply that there is also an edge from $j$ to $i$.
\vspace{1cm}

The lattice fermion has common properties with the spectral graph we have introduced here.
In the next sections, we will show that the lattice fermion can be identified as the spectral graph.


\section{Naive fermion}
\label{sec:naive}

In this section we discuss the Dirac operator matrix of Naive fermion in terms of SGT.
Through this example, we will find a Dirac operator of lattice fermions is identified as a matrix of the spectral graph corresponding to the lattice fermion.
We also show how to find the number of zero modes (the number of fermion species) by use of discrete Fourier transformation (DFT). This technique can be applied to any kind of matrices arising from the non-regular lattices.
By use of DFT, we will correctly derive the sixteen fermion species in the naive fermion.


\subsection{Dirac matrix of naive fermion}
\label{subsec:Dmat_naive}
	The lattice naive fermion action in four dimensions is
	\begin{equation}
		S = \sum_{n} \sum_{\mu=1}^{4} \barpsi_{n} \gam{\mu} \Dirac{\mu} \psi_{n}\,,
	\end{equation}
	where 
	$D_{\mu} \equiv (T_{+\mu} - T_{-\mu})/2$ with $T_{\pm\mu}\psi_{n} = U_{n,\pm\hmu}\psi_{n\pm\hmu}$ and $\hmu$ is a unit vector. In a free theory, we just set $\U{\pm\hmu} = \bm{1}$. The sum $\sum_{n}$ is the summation over lattice site $n = (n_{1}, n_{2}, n_{3}, n_{4})$ and those intervals are $1 \leq n_{\mu} \leq N$. Note that the spacetime where the fermion is defined is a four-dimensional torus because we usually impose periodic boundary conditions in each direction. 
	To derive the corresponding matrix of the lattice Dirac operator in a free theory, we introduce a vector of fermion fields. Namely, a vector $\psi$ is defined $\psi = \sum_{n} \psi_{n} \bm{e}_{n}$ where $\bm{e}_{n}$ are standard basis which satisfy orthonormal $\bm{e}_{n'}\bm{e}_{n} = \delta_{n'n} \equiv \prod_{i=1}^{4}\delta_{n'_{i}n_{i}}$. $\delta_{kl}$ is the Kronecker delta. Here we specify that the order of components $\psi_{n}$ in the vector is	$(1,1,1,1) \to \cdots \to (N,1,1,1) \to (1,2,1,1) \to \cdots \to (N,N,1,1) \to (1,1,2,1) \to \cdots \to (N,N,N,N) $ in descending order. Namely,
	\begin{equation}
		\psi = \bpmat
			\psi_{(1,1,1,1)} \\
			\vdots \\
			\psi_{(N,1,1,1)} \\
			\psi_{(1,2,1,1,1)} \\
			\vdots \\
			\psi_{(N,N,1,1)} \\
			\psi_{(1,1,2,1)} \\
			\vdots\\
			\psi_{(N,N,N,N)} \\
		\epmat
	\end{equation}
	in term of the vector.
	Thus, the action of naive fermion can be rewritten as $S = \barpsi \calD \psi = \sum_{m} \sum_{n} \barpsi_{m} \calD_{mn} \psi_{n}$ where $\calD$ is Dirac matrix having $\calD_{mn}$ as $(m,n)$ component. 
For later use, we now introduce the tensor-product representation:
By the tensor product (or Kronecker product), this matrix can be represented as
	\begin{equation}
	\begin{split}
		\calD
		&= \ide 	\ot \ide 	\ot \ide 	\ot \Pmat 	\ot \gam{1} \\
		&+ \ide 	\ot \ide 	\ot \Pmat	\ot \ide	\ot \gam{2} \\
		&+ \ide	\ot \Pmat	\ot \ide	\ot \ide	\ot \gam{3} \\
		&+ \Pmat	\ot \ide	\ot \ide	\ot \ide	\ot \gam{4}
	\end{split}
	\end{equation}
	where $\ide$ is an identity matrix of order $N$ and $\Pmat$ is a square matrix of order $N$ \cite{Kaveh2005,Sabidussi1959,Aurenhammer1992}. The components of the matrix $P_{N}$ is defined as $(\Pmat)_{ab} \equiv (E_{ab} - E_{ab}^{-1})/2$
where $E_{ab} \equiv \sum_{i=1}^{N-1} \delta_{ai} \delta_{i+1\, b} + \delta_{aN} \delta_{1b}$ and $E_{ab}^{-1} \equiv \delta_{a1}\delta_{Nb} + \sum_{i=2}^{N} \delta_{ai} \delta_{i-1\, b}$ for $a,b = 1,2,\cdots,N$. 
$\Pmat$ is explicitly written as
	\begin{equation}
	\label{eq:pmat}
		\Pmat
		= \frac{1}{2} \bpmat 
			0	&1		&0		&		&0		&0		&-1 \\
			-1	&0		&1		&\cdots	&0		&0		&0 \\
			0	&-1		&0		&		&0		&0		&0 \\
				&\vdots	&		&\ddots	&		&\vdots	& \\
			0	&0		&0		&		&0		&1		&0 \\
			0	&0		&0		&\cdots	&-1		&0		&1 \\
			1	&0		&0		&		&0		&-1		&0
		\epmat
	\end{equation}
	and $E$ and $E^{\dagger}$ are similarly represented as
	\begin{equation}
		E = \bpmat
			0	&1		&0	&		&0		&0		&0\\
			0	&0		&1	&\cdots	&0		&0		&0\\
			0	&0		&0	&		&0		&0		&0\\
				&\vdots	&	&\ddots	&		&\vdots	&\\
			0	&0		&0	&		&0		&1		&0\\
			0	&0		&0	&\cdots	&0		&0		&1\\
			1	&0		&0	&		&0		&0		&0
		\epmat, \quad
		E^{\dagger} = \bpmat
			0	&0		&0	&		&0		&0		&1\\
			1	&0		&0	&\cdots	&0		&0		&0\\
			0	&1		&0	&		&0		&0		&0\\
				&\vdots	&	&\ddots	&		&\vdots	&\\
			0	&0		&0	&		&0		&0		&0\\
			0	&0		&0	&\cdots	&1		&0		&0\\
			0	&0		&0	&		&0		&1		&0
		\epmat.
	\end{equation}
	Note that the matrix $\Pmat$ is the matrix representing the periodic boundary condition of one direction.
	This matrix is the circulant matrix.

\subsection{Graphs corresponding to naive fermion}
\label{subsec:graph_naive}
	
In this subsection, based on the matrix representations we have obtained in the previous subsection, we see how the spectral graphs correspond to the naive lattice fermions on low- and high-dimensional lattices. 
Although in the cases of non-torus lattice fermions we obtain the Dirac matrix from the spectral graph,
we now derive ``the graph from the matrix" conversely.	

\begin{figure}[htbp]
	\centering
		\includegraphics[clip,keepaspectratio, scale=.7]{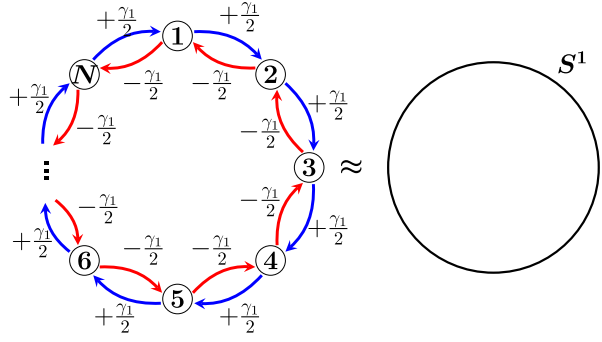}
\caption{This graph corresponds to the Dirac matrix of one dimensional naive fermion with periodic boundary condition. This graph schematically shows a circle $S^{1}$.}
		\label{graph:naive_1d}
	\end{figure}

Firstly, we consider the one dimensional lattice. 
In this case, the Dirac matrix of the free naive fermion is 
	\begin{equation}
		\calD^{\mathrm{1d}} = P_{N} \ot \gamma_{1}.
	\end{equation}
Now, the graph corresponding to this Dirac matrix is depicted in Fig.~\ref{graph:naive_1d}. 
This graph schematically represents a circle $S^{1}$, where we impose a periodic boundary condition.
This is called a circulant graph in graph theory.
	
\begin{figure}[htbp]
\centering
	\includegraphics[clip,keepaspectratio, scale=.5]{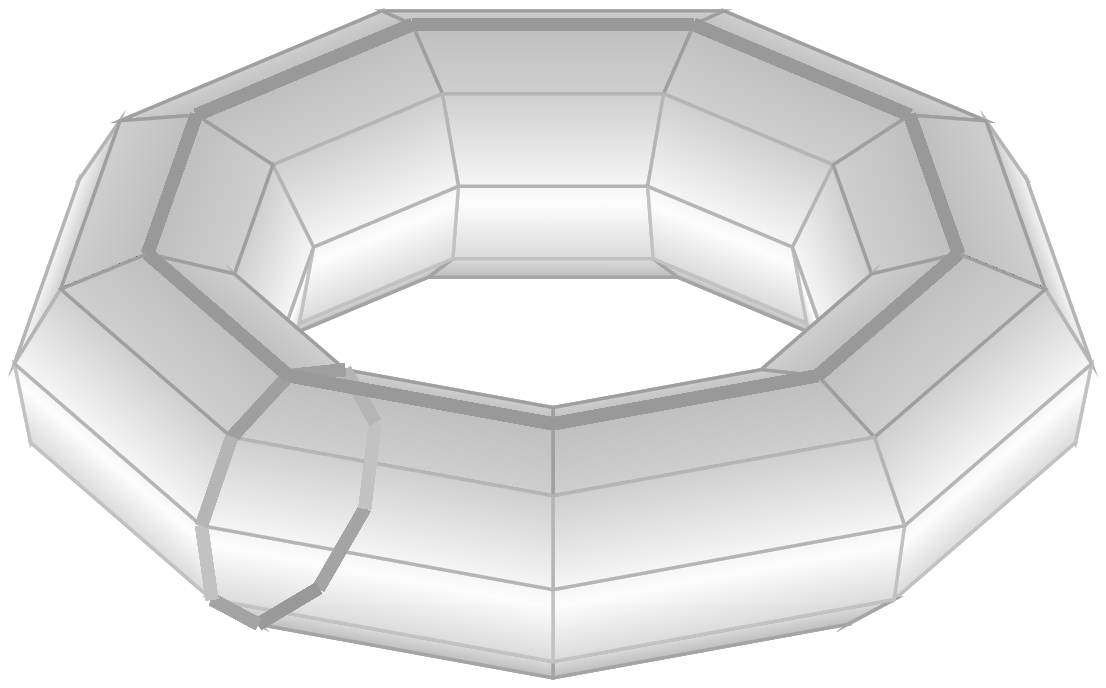}
\caption{
This graph corresponds to the Dirac matrix of two-dimensional naive fermion with periodic boundary condition. This graph schematically represents the two-dimensional torus $T^{2}$. 
The weight for each edge from $n$ to $n+\hat{\mu}$ is $+\gam{\mu}/2$ and the one for each edge from $n$ to $n-\hat{\mu}$ is $-\gam{\mu}/2$.
}
\label{graph:naive_2d}
\end{figure}

In the case of two-dimensional naive fermion, the Dirac matrix is 
\begin{equation}
	D^{\mathrm{2d}} = \ide \ot \Pmat \ot \gam{1} + \Pmat \ot \ide \ot \gam{2}\,,
\end{equation}
and the graph, who schematically represents two-dimensional torus $T^{2}$, is expressed as Fig.~\ref{graph:naive_2d}, where each direction is independent, leading to $S^{1} \times S^{1} \sim T^{2}$.
There are both an edge with the positive weight and an edge with the negative weight between each vertex. 
It is notable that the weight for edges from $n$ to $n+\hat{\mu}$ is $+\gam{\mu}$ and the one for edges from $n$ to $n-\hat{\mu}$ is $-\gam{\mu}$.

In three and four dimensions, the procedures are parallel. The corresponding graphs schematically represent $S^{1} \times S^{1} \times S^{1} \sim T^{3}$ and $S^{1} \times S^{1} \times S^{1} \times S^{1} \sim T^{4}$, 
respecctively.
For the cases of non-torus lattice fermions, we conversely obtain the Dirac matrix from the spectral graph.

\subsection{Diagonalization of Dirac matrix $\calD$}

	In this subsection, we discuss how to diagonalize Dirac matrix $\calD$ and find the number of zero eigenvalues for a free four-dimensional naive fermion. Firstly, we will diagonalize $E$ in order to diagonalize the whole Dirac matrix. We introduce the discrete Fourier transform (DFT) matrix $X$, which is consisted of $X_{jk} \equiv \xi^{(k-1)(j-1)}/ \sqrt{N} $ with $\xi = e^{-\frac{2\pi i}{N}}$ for $j,k = 1,2,\cdots,N$. It is clear that DFT matrix is unitary. Then, $E$ and $E^{\dagger}$ satisfy $EX = X\Lambda$ and $E^{\dagger} X = \Lambda^{\dagger} X$, where $\Lambda$ is a diagonal matrix as $\Lambda = \diag{1,\xi, \xi^{2}, \cdots, \xi^{N-1}}$. Since $P_{N} = E - E^{\dagger}$, the diagonalized $P_{N}$ is given as
	\begin{equation}
		P_{N}X 
		= i\, \diag{0,\, \sin \frac{2\pi}{N},\, \sin \frac{4\pi}{N},\, \cdots,\, \sin \frac{2(N-1)\pi}{N}} X
		\equiv \Lambda_{P_{N}} X.
	\end{equation}
	There exist two zero eigenvalues in $\Lambda_{P_{N}}$ at most when $N$ is an even number. Now, the Dirac matrix $\calD$ can be diagonalized as
	\begin{equation}
	\begin{split}
	\label{diag:Dirac}
		\Uni^{\dagger} \calD\, \Uni
		&= \ide 	\ot \ide 	\ot \ide 	\ot \Lambda_{\Pmat} 	\ot \gam{1} \\
		&+ \ide 	\ot \ide 	\ot \Lambda_{\Pmat}	\ot \ide	\ot \gam{2} \\
		&+ \ide	\ot \Lambda_{\Pmat}	\ot \ide	\ot \ide	\ot \gam{3} \\
		&+ \Lambda_{\Pmat}	\ot \ide	\ot \ide	\ot \ide	\ot \gam{4}
	\end{split}
	\end{equation}
	where $\Uni$ is a unitary matrix defined as $\Uni = \bigotimes_{\mu=1}^{4} X \ot \bm{1}_{4}$. 
	It is easy to see the number zero eigenvalues of the Dirac matrix by introducing eigenvectors of $E$. We denote eigenvectors of $E$ as $\ket{k}$ for $k= 1,2,\cdots, N$. They satisfy $E \ket{k} = \xi^{k-1} \ket{k}$ and $\braket{k' | k} = \delta_{k'k}$. Then, the unitary matrix $\Uni$ is written as $\Uni = \sum_{k_{4},k_{3},k_{2},k_{1}} \ket{k_{4},k_{3},k_{2},k_{1}}\bra{k_{4},k_{3},k_{2},k_{1}} \ot \bm{1}_{4}$, where eigenvectors $\ket{k_{4},k_{3},k_{2},k_{1}}$ mean $\ket{k_{4}} \ot \ket{k_{3}} \ot \ket{k_{2}} \ot \ket{k_{1}}$. Thus the diagonalization of the Dirac matrix is expressed as
	\begin{equation}
	\label{eigen_eq}
		\Uni^{\dagger} \calD\, \Uni
		= \sum_{k_{4},k_{3},k_{2},k_{1}} \left[ 
			i \sum_{\mu=1}^{4} \sin \left( \frac{2\pi(k_{\mu} - 1)}{N} \right) \gam{\mu}
		\right] \ket{k_{4},k_{3},k_{2},k_{1}}\bra{k_{4},k_{3},k_{2},k_{1}}.
	\end{equation}
	If $\Uni^{\dagger} \calD\, \Uni$ has zero eigenvalues, Eq.~(\ref{eigen_eq}) must satisfy an equation below
	\begin{equation}
	\label{period_eq}
		\sum_{\mu=1}^{4} \sin \left( \frac{2\pi(k_{\mu} - 1)}{N} \right) \gam{\mu} = \bm{0}.
	\end{equation}
	However, since $\gamma$ matrices are linearly independent, the coefficient of each $\gamma$ matrices must be zero.
	Finally, the conditions for the diagonalized Dirac matrix to have zero eigenvalues are
	\begin{equation}
	\label{condition:naive}
		\sin \left( \frac{2\pi(k_{\mu} - 1)}{N} \right) =0.
	\end{equation}
	The solutions of Eq.~(\ref{condition:naive}) are $k_{\mu} = 1, \frac{N}{2}+1$. 
	In this case, we do not need to take a continuum and a thermodynamic limit as long as we take an even number as $N$.
	Therefore, the number of zero eigenvalues are $2^{4}$ since the number of zero eigenvalues is equal to the number of combination of the solutions in four dimensions. This result is consistent with the well-known number of doublers in four-dimensional naive fermion.
	
In the end of this section, we comment on the meaning of $k_{\mu}$.
This $k_{\mu}$ can be interpreted as a ``shifted momentum" in this case.
However, it is not necessarily equivalent to the momentum since it can be defined also on the lattice in which the momentum cannot be defined.
Thus, $k_{\mu}$ should be simply interpreted as a variable for the diagonalization in general.


\section{Wilson fermion}
\label{sec:Wilson}
\subsection{Dirac matrix of Wilson fermion}
In this section we discuss the Dirac operator matrix of Wilson fermion in terms of SGT.
The lattice Wilson fermion action in four dimensions is expressed as
	\begin{equation}
		S_{W} 
		= \sum_{n} \sum_{\mu} \bar{\psi}_{n} \gamma_{\mu} D_{\mu} \psi_{n}
		+ m \sum_{n} \bar{\psi}_{n} \psi_{n} \\
		+ r \sum_{n} \sum_{\mu} \bar{\psi}_{n} \left( 1-C_{\mu} \right) \psi_{n}	\,,
	\end{equation}
	with $C_{\mu} \equiv (T_{+\mu} + T_{-\mu})/2$. $m$ is a mass parameter and $r$ is a Wilson-fermion parameter. The periodic boundary condition is imposed on each direction and those intervals are given as $1\leq n_{\mu}\leq N$.
	By investigating the action in a manner similar to Sec.~\ref{sec:naive}, it is written as $S_{W} = \barpsi \calD_{W} \psi \equiv \barpsi \left( \calD + \calM_{W} \right) \psi$, where $\calD_{W}$ is the Dirac matrix of Wilson fermion and 
$\mathcal{M}_{W}$ is mass matrix consisting of the Wilson term. The mass matrix $\calM_{W}$ is represented by use of tensor products as
\begin{equation}
\begin{split}
	\mathcal{M}_{W}
	&= m\cdot \ide \otimes \ide \otimes \ide \otimes \ide \ot \idef \\
	&+r\cdot \bigl\{
			\ide		\otimes	\ide		\otimes	\ide		\otimes	M_{W}		\ot	\idef +
			\ide		\otimes	\ide		\otimes	M_{W}	\otimes	\ide 			\ot	\idef\\
	&+
			\ide		\otimes	M_{W}	\otimes	\ide		\otimes	\ide 			\ot	\idef +
			M_{W}	\otimes	\ide		\otimes	\ide		\otimes	\ide 			\ot	\idef
	\bigr\}
\end{split}
\end{equation}
where $M_{W} = \ide - (E + E^{\dagger})/2$ and the product $\cdot$ stands for scalar product.

\subsection{Graphs corresponding to Wilson fermion}
\begin{figure}[htbp]
	\centering
		\includegraphics[clip,keepaspectratio, scale=.7]{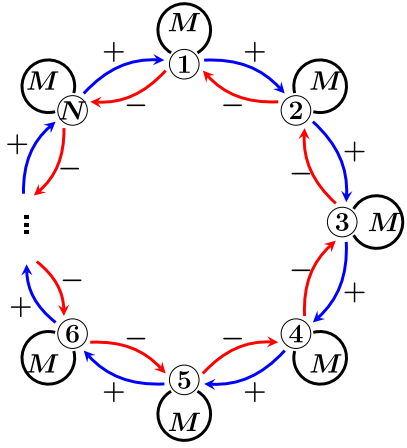}
\caption{
This digraph corresponds to the Dirac matrix of one-dimensional Wilson fermion. The weight of the loop is $M = m+r$. The weight $\pm$ for each edge stand for $\pm (\gam{1} \mp r\idef)/2$.
}
\label{graph:Wf_1d}
\end{figure}

As with the case in Sec.~\ref{subsec:graph_naive}, we begin with the one-dimensional lattice. 
For one-dimensional Wilson fermion, the Dirac matrix is given by
\begin{equation}
	D_{W}^{\mathrm{1d}} = \Pmat \ot \gam{1} + m \ide \ot \idef + r M_{W} \ot \idef ,
\end{equation}
and the graph corresponding to this matrix is depicted in Fig.~\ref{graph:Wf_1d}. 
This graph again represents the circle $S^{1}$.
However, the difference from the case of naive fermions in Fig.~\ref{graph:naive_2d} is 
that Fig.~\ref{graph:Wf_1d} has a loop with the weight $M=m+r$ for each vertex. 

\begin{figure}[htbp]
\centering
	\includegraphics[clip,keepaspectratio, scale=.5]{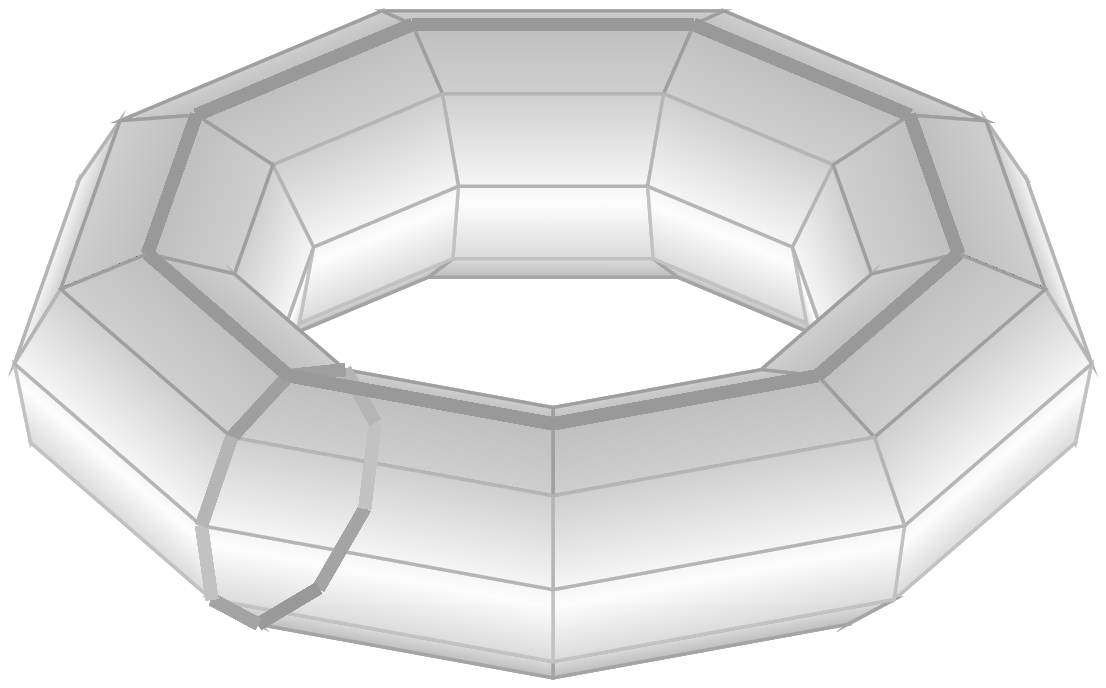}
\caption{
	This digraph corresponds to the Dirac matrix in two-dimensional Wilson fermion. The shape of this graph is topologically equivalent to two-dimensional torus $T^{2}$. The weight for each edge from $n$ to $n+\hat{\mu}$ is $+(\gam{\mu}-r\idef)/2$ and the one for each edge from $n$ to $n-\hat{\mu}$ is $-(\gam{\mu}+r\idef)/2$. The weight for each loop is $M = m+\sum_{\mu}r$.
}
\label{graph:Wilson_2d}
\end{figure}

In two dimensions, the Dirac matrix is 
\begin{equation}
\begin{split}
	D_{W}^{\mathrm{2d}} &= \ide \ot \Pmat \ot \gam{1} + \Pmat \ot \ide \ot \gam{2}  \\
	&+ m \left( \ide \ot \ide \ot \idef \right) + r \left( \ide \ot M_{W} + M_{W} \ot \ide \right) \ot \idef
\end{split}
\end{equation}
and the graph corresponding to this matrix is given by Fig.~\ref{graph:Wilson_2d}. The shape of the graph is topologically a two-dimensional torus $T^{2}$ however there is a loop with the weight $M = m+\sum_{\mu}r$ for each vertex. Furthermore, the weight for each edge from $n$ to $n+\hat{1}$ is $+(\gam{1} - r\idef)/2$ and the one for each edge from $n$ to $n-\hat{1}$ is $-(\gam{1} + r\idef)/2$. In other direction $\mu = 2$, we just replace $\gam{1}$ by $\gam{2}$.

The shape of the graph corresponding to Wilson fermion in four dimensions is again topologically a four-dimensional torus $T^{4}$. Furthermore, there is a loop for each vertex, which has the weight $M = m + \sum_{\mu}r$. The weight for each edge from $n$ to $n+\hat{\mu}$ is $+(\gam{\mu} - r\idef)/2$ while the wight for each edge from $n$ to $n-\hat{\mu}$ is $-(\gam{\mu} + r\idef)/2$.

\subsection{Diagonalization of Dirac matrix $\calD_{W}$}

Firstly, we will diagonalize the mass matrix $\calM_{W}$ so as to diagonalize the whole Dirac matrix $\calD_{W}$. As with the discussion of naive fermion, we use the DFT matrix $X$. By using the matrix $X$ we can diagonalize $M_{W}$ as
	\begin{equation}
	M_{W}X = \diag{0,\, 1-\cos \frac{2\pi}{N},\, 1-\cos \frac{4\pi}{N},\, \cdots,\, 1-\cos \frac{2(N-1)\pi}{N}} X
		\equiv \Lambda_{M_{W}} X.
	\end{equation}
	There is only a single zero eigenvalue in $\Lambda_{M_{W}}$ at most if $N$ is even number.
	As a result the diagonalization is expressed as
	\begin{equation}
	\begin{split}
		\Uni^{\dagger} \calM_{W}\, \Uni
		&= m \cdot \ide \ot \ide \ot \ide \ot \ide \ot \idef \\
		&+r\cdot \bigl\{
			\ide \ot \ide \ot \ide \ot \Lambda_{M_{W}} \ot \idef +
			\ide \ot \ide \ot \Lambda_{M_{W}} \ot \ide \ot \idef \\
		&+
			\ide \ot \Lambda_{M_{W}} \ot \ide \ot \ide \ot \idef +
			\Lambda_{M_{W}} \ot \ide \ot \ide \ot \ide \ot \idef 
		\bigr\}
	\end{split}
	\end{equation}
	where $\Uni = \bigotimes_{\mu=1}^{4} X \ot \bm{1}_{4}$. Since the Dirac matrix of naive fermion $\calD$ can be diagonalized with the unitary matrix $U$, the Dirac matrix of Wilson fermion can be diagonalized as $\Uni^{\dagger} \calD_{W}\, \Uni = \Uni^{\dagger}\calD\, \Uni + \Uni^{\dagger} \calM_{W}\, \Uni$. We introduce the eigenvectors $\ket{k}$ of $E$ to see the number of zero eigenvalues in the diagonalized Dirac matrix $\Uni^{\dagger} \calD_{W}\, \Uni$. The diagonalized mass matrix $\Uni^{\dagger} \calM_{W}\, \Uni$ is written as
	\begin{equation}
	\begin{split}
		\Uni^{\dagger} \calM_{W}\, \Uni
		= &\sum_{k_{4},k_{3},k_{2},k_{1}} \left[
			m\,\idef + r \sum_{\mu=1}^{4} \left\{ 1 - \cos \left( \frac{2\pi(k_{\mu}-1)}{N} \right) \right\}  \idef
		\right] \\
		&\times \ketk \brak.
	\end{split}
	\end{equation}
	Thus the diagonalized Dirac matrix $\Uni^{\dagger} \calD_{W}\, \Uni$ is
	\begin{equation}
	\label{eigen:Dirac_W}
	\begin{split}
		\Uni^{\dagger} \calD_{W}\, \Uni
		= &\sumk \left[ i\sum_{\mu} \sin \left( \frac{2\pi(k_{\mu}-1)}{N} \right) \gam{\mu}
			+ r \sum_{\mu} \left\{ \frac{m}{4r} + 1 - \cos \left( \frac{2\pi(k_{\mu}-1)}{N} \right) \right\} \idef \right] \\
		&\times \ketk \brak.
	\end{split}
	\end{equation}
	If the diagonalized Dirac matrix $\Uni^{\dagger} \calD_{W}\, \Uni$ has zero eigenvalues, Eq.~(\ref{eigen:Dirac_W}) must satisfy a equation
	\begin{equation}
		i\sum_{\mu} \sin \left( \frac{2\pi(k_{\mu}-1)}{N} \right) \gam{\mu}
			+ r\sum_{\mu} \left\{ \frac{m}{4r} + 1 - \cos \left( \frac{2\pi(k_{\mu}-1)}{N} \right) \right\} \idef = \bm{0}
	\end{equation}
	Since $\gamma$ matrices and $\idef$ are linearly independent, 
	the conditions for the diagonalized Dirac matrix to have zero eigenvalues are given as
	\begin{gather}
	\label{condition:Wilson1}
		\sin \left( \frac{2\pi(k_{\mu} - 1)}{N} \right) =0 \\
	\label{condition:Wilson2}
		m+4r - r\sum_{\mu} \cos \left( \frac{2\pi(k_{\mu}-1)}{N} \right) = 0\,.
	\end{gather}
	The solutions of Eq. (\ref{condition:Wilson1}) are $k_{\mu} = 1, \frac{N}{2} + 1$. 
	These solutions are classified by Eq.~(\ref{condition:Wilson2}) as
	\begin{equation}
		M(k) = 
		\begin{cases}
			m		&\mbox{any $k_{\mu}=1$ in $k$.}\\
			m+2r		&\mbox{one $k_{\mu}=1+N/2$ otherwise $k_{\mu}=1$ in $k$.}\\
			m+4r		&\mbox{two $k_{\mu}=1+N/2$ otherwise $k_{\mu}=1$ in $k$.} \\
			m+6r		&\mbox{three $k_{\mu} = 1+N/2$ otherwise $k_{\mu}=1$ in $k$.}\\
			m+8r		&\mbox{any $k_{\mu}=1+N/2$ in $k$.}
		\end{cases}
	\end{equation}
	where we define $M(k) \equiv m+4r - r\sum_{\mu} \cos \left( \frac{2\pi(k_{\mu}-1)}{N} \right)$ for $k = (k_{1},k_{2},k_{3},k_{4})$ in Eq.~(\ref{condition:Wilson2}). Hence, the number of zero eigenvalues of Dirac matrix $\calD_{W}$ depends on mass parameter $m$ and Wilson-fermion parameter $r$. If we set $m=0$ and $r=1$, the number of zero eigenvalues is just one. In this case, we do not need to take a continuum and a thermodynamic limit as long as we take an even number as $N$. This result is consistent with the known result of Wilson fermion.

\section{Domain-wall fermion}
\label{sec:DW}
\subsection{Dirac matrix and mass matrix of domain-wall fermion}
In this section we discuss the Dirac operator matrix of domain-wall fermions in terms of SGT.
The domain-wall fermion action in five dimensions (4+1 dimensions) is
	\begin{equation}
	\begin{split}
		S_{DW} 
		= &\sum_{n,s} \bar{\psi}_{n,s}
			\left[ \sum_{\mu} \gamma_{\mu} D_{\mu} \psi_{n,s} + \gamma_{5} D_{s} \psi_{n,s} \right]
		- M_{0} \sum_{n,s} \bar{\psi}_{n,s} \psi_{n,s} \\
		&+ \sum_{n,s} \bar{\psi}_{n,s} \left[ \sum_{\mu} \left( 1-C_{\mu} \right) \psi_{n,s} + \left( 1 - C_{s} \right) \psi_{n,s} \right]
	\end{split}
	\end{equation}
	where $D_{s}\psi_{n,s} \equiv (\psi_{n,s+1} - \psi_{n,s-1})/2,\ C_{s}\psi_{n,s} \equiv (\psi_{n,s+1} + \psi_{n,s-1})/2$ and $M_{0}$ is a mass parameter. The sum $\sum_{n,s}$ is the summation over five-dimensional lattice sites $n,s = (n_{1}, n_{2}, n_{3}, n_{4}; s)$. Note that there are two differences regarding boundary conditions and dynamical (or non-dynamical) variables between the fifth direction and other directions:
	\begin{itemize}
	\item
	On the fifth direction $s$, we impose the Dirichlet boundary condition, $\psi_{n,0} = \psi_{n,N_{s}+1} = 0$ , and its interval is $1 \leq s \leq N_{s}$.
	We impose a periodic boundary condition on other directions $n$ and those intervals are $1 \leq n_{\mu} \leq N$.
	 \item
	 The link variables in the fifth direction are not dynamical while those in other directions can be dynamical.
	 \end{itemize}
	 Based on a similar discussion in Sec.~\ref{sec:naive}, the action of a free domain-wall fermion can be rewritten as $S_{DW} =\sum_{s,t} \barpsi_{s} \left[ \calD_{DW} + \calM_{DW} \right]_{st} \psi_{t}$ for $s,t = 1,2,\cdots, N_{s}$, where $\calD_{DW}$ is the Dirac matrix of domain-wall fermion and $\calM_{DW}$ is a mass matrix of domain-wall fermion from four-dimensional viewpoints. 
	 The Dirac matrix $\calD_{DW}$ is 
	\begin{equation}
	\begin{split}
		\left[ \calD_{DW} \right]_{st}
		= &\delta_{st} \cdot \bigl(\, \ide \ot \ide \ot \ide \ot \Pmat \ot \gam{1} 
		+ \ide \ot \ide \ot \Pmat \ot \ide \ot \gam{2} \\
		&+ \ide \ot \Pmat \ot \ide \ot \ide \ot \gam{3} 
		+ \Pmat \ot \ide \ot \ide \ot \ide \ot \gam{4} \,\bigr)
	\end{split}
	\end{equation}
	and the mass matrix $\calM_{DW}$ is 
	\begin{equation}
	\begin{split}
		\left[ \calM_{DW} \right]_{st}
		= &\frac{1}{2} \Delta_{st}^{(-)} \cdot \ide \ot \ide \ot \ide \ot \ide \ot \gam{5} \\
		&+ \delta_{st} \bigl(\, -M_{0}\cdot \ide \ot \ide \ot \ide \ot \ide \ot \idef \\
		&+ \ide \ot \ide \ot \ide \ot M_{W} \ot \idef 
			+ \ide \ot \ide \ot M_{W} \ot \ide \ot \idef \\
		&+ \ide \ot M_{W} \ot \ide \ot \ide \ot \idef 
			+ M_{W} \ot \ide \ot \ide \ot \ide \ot \idef \,\bigr) \\
		&+ \frac{1}{2} \left( 2\delta_{st} - \Delta_{st}^{(+)} \right) \cdot \ide \ot \ide \ot \ide \ot \ide \ot \idef
	\end{split}
	\end{equation}
	where $\Delta_{st}^{(\pm)} = \sum_{i=1}^{N-1} \delta_{si} \delta_{i+1\,t} \pm \sum_{l=2}^{N} \delta_{sl} \delta_{l-1\,t}$.
	Note that we specify the order of components $(\psi_{n})_{s}$ in the vector $\psi_{s}$ as $(1,1,1,1;s) \to \cdots \to (N,1,1,1;s) \to (1,2,1,1;s) \to \cdots \to (N,N,1,1;s) \to (1,1,2,1;s) \to \cdots \to (N,N,N,N;s) $ in a descending order.
	For simplicity, we introduce the chiral projection $\proj_{R}$ and $\proj_{L}$ for the mass matrix $\calM_{DW}$. The matrix $\calM_{DW}$ can then be rewritten as
	\begin{gather}
		\left[ \calM_{DW} \right]_{st} = \calM_{st}^{(L)} \ot \proj_{L} + \calM_{st}^{(R)} \ot \proj_{R}\\
	\begin{split}
		\calM_{st}^{(L)} 
		= &\delta_{st} \bigl\{\, \left( 1-M_{0} \right) \cdot \ide \ot \ide \ot \ide \ot \ide \\
		&+ \ide \ot \ide \ot \ide \ot M_{W} + \ide \ot \ide \ot M_{W} \ot \ide \\
		&+ \ide \ot M_{W} \ot \ide \ot \ide + M_{W} \ot \ide \ot \ide \ot \ide \,\bigr\} \\
		&- \sum_{i=1}^{N-1} \delta_{si} \delta_{i+1\,t} \cdot \ide \ot \ide \ot \ide \ot \ide 
	\end{split}\\
	\begin{split}
		\calM_{st}^{(R)} 
		= &\delta_{st} \bigl\{\, \left( 1-M_{0} \right) \cdot \ide \ot \ide \ot \ide \ot \ide \\
		&+ \ide \ot \ide \ot \ide \ot M_{W} + \ide \ot \ide \ot M_{W} \ot \ide \\
		&+ \ide \ot M_{W} \ot \ide \ot \ide + M_{W} \ot \ide \ot \ide \ot \ide \,\bigr\} \\
		&- \sum_{l=2}^{N} \delta_{sl} \delta_{l-1\,t} \cdot \ide \ot \ide \ot \ide \ot \ide 
	\end{split}
	\end{gather}
	where $\proj_{R} \equiv (\idef + \gam{5})/2$ is a projection matrix into right-handed components and $\proj_{L} \equiv (\idef - \gam{5})/2$ is into left-handed components.

\subsection{Graphs corresponding to domain-wall fermion}
\begin{figure}[htbp]
\centering
	\includegraphics[clip,keepaspectratio, scale=.7]{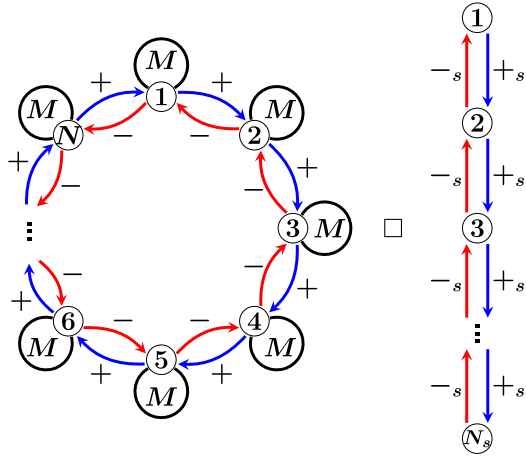}
\caption{This digraph corresponds to the Dirac matrix and mass matrix in the two-dimensional domain-wall fermion. The left graph is equivalent to one-dimensional Wilson fermion. On the other hand, the right graph corresponds to the extra direction $s$. The weight $+_{s}$ means $+(\gam{3}-\idef)/2$ and $-_{s}$ means $-(\gam{3} + \idef)/2$. {\Large $\Box$} stands for the Cartesian product.}
\label{graph:DW}
\end{figure}

We now discuss a graph corresponding to domain-wall fermion in two dimensions ($1+1$ dimensions).
For two-dimensional domain-wall fermion, the Dirac matrix is given as
\begin{equation}
	\left[ \calD_{DW}^{\mathrm{1d}} \right]_{st} = \delta_{st} \Pmat \ot \gam{1}
\end{equation}
and the mass matrix is
\begin{equation}
\begin{split}
	\left[ \calM_{DW}^{\mathrm{1d}} \right]_{st} 
	&= \frac{1}{2} \Delta_{st}^{(-)} \cdot \ide \ot \gam{3} 
		+ \delta_{st} \left( -M_{0} \ide \ot \idef + M_{W} \ot \idef \right) 
		+ \frac{1}{2} \left( 2\delta_{st} - \Delta_{st}^{(+)} \right) \ide \ot \idef \\
	&= {\calM_{st}^{\mathrm{1d}}}^{(L)} \ot \proj_{L}^{\mathrm{1d}} 
		+ {\calM_{st}^{\mathrm{1d}}}^{(R)} \ot \proj_{R}^{\mathrm{1d}}
\end{split}
\end{equation}
where
\begin{gather}
	{\calM_{st}^{\mathrm{1d}}}^{(L)}
	= \delta_{st} \left\{ \left( 1-M_{0} \right) \ide + M_{W} \right\} - \sum_{i=1}^{N-1} \delta_{si} \delta_{i+1\,t} \ide \\
	{\calM_{st}^{\mathrm{1d}}}^{(R)}
	= \delta_{st} \left\{ \left( 1-M_{0} \right) \ide + M_{W} \right\} - \sum_{l=2}^{N} \delta_{sl} \delta_{l-1\,t} \ide \\
	\proj_{L}^{\mathrm{1d}} 
	= \frac{1}{2} \left( \idef - \gam{3} \right), \quad 
	\proj_{R}^{\mathrm{1d}}
	= \frac{1}{2} \left( \idef + \gam{3} \right) \,.
\end{gather}
We adopt $\gam{3}$ as the extra dimensional gamma matrix.
A graph corresponding to the Dirac matrices is Fig.~\ref{graph:DW}. 
The schematicl shape of this graph is paetially similar to the graph in one-dimensional Wilson fermion. 
There is also a loop with the weight $M=2-M_{0}$ for each vertex. 
Each edge from $n$ to $n+\hat{1}$ has the weight $+(\gam{1} - \idef)/2$ and the one from $n$ to $n-\hat{1}$ has the weight $-(\gam{1} + \idef)/2$.
However, there is another direction $s$ and each edge from $s$ to $s\pm1$ has the weight $\pm (\gam{5} \mp \idef)/2$. 
The two graphs corresponding to the direction $\mu=1$ and in the extra direction $s$ is producted by Cartesian product $\Box$. 
It is notable that a graph in the direction $\mu=1$ is equivalent to a graph in one-dimensional Wilson fermion.
	
For the five-dimensional domain-wall fermion, a graph also consists of the two graphs, a graph corresponding to the four dimensional Wilson fermion and a graph in the extra direction $s$, which are unified by Cartesian product.

\subsection{Diagonalization of Dirac matrix $\calD_{DW}$ and mass matrix $\calM_{DW}$}

Firstly, We diagonalize the Dirac matrix $\calD_{DW}$ in a parallel way to Sec.~\ref{sec:naive}. 
By using the unitary matrix $\Uni$, the diagonalized Dirac matrix is given as
	\begin{equation}
		\bigl[ \Uni^{\dagger}\calD_{DW} \,\Uni \bigr]_{st}
		= \delta_{st} \sumk \left[ i \sum_{\mu=1}^{4} \sin \left( \frac{2\pi(k_{\mu} - 1)}{N} \right) \gam{\mu} \right]
			\ketk \brak
	\end{equation}
	for $k_{\mu} = 1,2, \cdots, N$. As a result, the solutions of $k_{\mu}$  for the Dirac matrix $\calD_{DW}$ to have zero eigenvalues are given by $k_{\mu} = 1,1+N/2$.

Secondly, we will diagonalize the mass matrix $\calM_{DW}$. 
The diagonalized mass matrix is
	\begin{equation}
	\label{eq:DW-mass}
	\begin{split}
		\bigl[\, \Uni^{\dagger} \calM_{DW}\, \Uni \,\bigr]_{st}
		=\ &\Uni^{\dagger} \left( \calM_{st}^{(L)} \ot \proj_{L} \right) \Uni 
			+ \Uni^{\dagger} \left( \calM_{st}^{(R)} \ot \proj_{R} \right) \Uni \\
		= &\sumk \left[ \left\{ W(k)\delta_{st} - \sum_{i=1}^{N-1} \delta_{si} \delta_{i+1\,t} \right\} \proj_{L}
			+  \left\{ W(k)\delta_{st} - \sum_{l=2}^{N} \delta_{sl} \delta_{l-1\,t} \right\} \proj_{R} \right] \\
		&\times \ketk \brak
	\end{split}
	\end{equation}
	where $W(k) \equiv 1-M_{0} + \sum_{\mu} \left\{ 1 - \cos \left( \frac{2\pi(k_{\mu}-1)}{N} \right) \right\}$. 
We now impose conditions, $|W(k)| < 1$
and $N_{s} = \infty$ so that mass matrix $\calM_{DW}$ has zero eigenvalues. 
We can determine the range of mass parameter $M_{0}$ from $|W(k)| < 1$.
Using the solutions of $k_{\mu}$, $W(k)$ can be classified as
	\begin{equation}
		W(k) = \begin{cases}
			1-M_{0} 	&\mbox{any $k_{\mu}=1$ in $k$.}\\
			3-M_{0}	&\mbox{one $k_{\mu}=1+N/2$ otherwise $k_{\mu}=1$ in $k$.}\\
			5-M_{0}	&\mbox{two $k_{\mu}=1+N/2$ otherwise $k_{\mu}=1$ in $k$.} \\
			7-M_{0}	&\mbox{three $k_{\mu} = 1+N/2$ otherwise $k_{\mu}=1$ in $k$.}\\
			9-M_{0}	&\mbox{any $k_{\mu}=1+N/2$ in $k$.}
		\end{cases}
	\end{equation}	
Therefore, the relation between the solutions of $k_{\mu}$, the range of mass parameter $M_{0}$, and the number of zero eigenvalues are given in Table.~\ref{table:DW}.
The result is consistent with the known results of the domain-wall fermion.
It indicates that our method can be applied to the lattice fermion with Dirichlet boundary conditions.

	\begin{table}[htbp]
	\caption{Classification of the number of zero eigenvalues in DW fermion}
	\label{table:DW}
	\centering
	\begin{tabular}{l c c}
	\hline
	the solutions of $k_{\mu}$&
		the range of $M_{0}$	&
			the number of zero eigenvalues \\
	\hline \hline
	any $k_{\mu} = 1$ &
		$0 < M_{0} < 2$ &
			1\\
	one $k_{\mu} = 1+N/2$ otherwise $k_{\mu} = 1$ &
		$2 < M_{0} < 4$ &
			4\\
	two $k_{\mu} = 1+N/2$ otherwise $k_{\mu} = 1$ &
		$4 < M_{0} < 6$ &
			6\\
	three $k_{\mu} = 1+N/2$ otherwise $k_{\mu} = 1$ &
		$6 < M_{0} < 8$ &
			4\\
	any $k_{\mu} = 1+N/2$ &
		$8 < M_{0} < 10$ &
			1\\
	\hline
	\end{tabular}
	\end{table}

\section{Lattice fermions on four dimensional hyperball}
\label{sec:lf_on_Bd}
In Sec. \ref{sec:naive},\ref{sec:Wilson},\ref{sec:DW}, we show that there are the spectral digraphs corresponding to the Dirac matrices of the naive fermion, Wilson fermion, and domain-wall fermion. 
Then, we find out the number of fermion species based on the discrete Fourier transformation. 

Our question is whether we can obtain the Dirac matrix when the spectral digraph corresponding to a lattice fermion is given.
In this section, we will construct the Dirac matrix from a digraph corresponding to a certain lattice fermion and derive the number of fermion species by use of DFT. 
Through this procedure, we can construct the Dirac matrix for a lattice fermion defined on complicated lattices, where the momentum cannot be defined. 
Then, we can derive the number of the species for fermions on such lattices.

From the next subsection we consider the discretized four-dimensional hyperball $B^{4}$.
This space is obtaining by imposing the Dirichlet boundary condition on each direction of four dimensions.


\subsection{Graphs corresponding to the Dirac matrix on hyperball}
\begin{figure}[htbp]
\centering
	\includegraphics[clip,keepaspectratio, scale=.8]{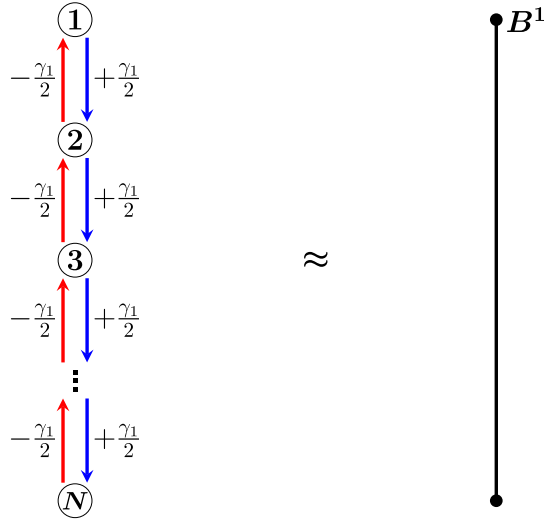}
\caption{This digraph corresponds to the lattice fermion on the discretized $B^{1}$.}
\label{graph:B1}
\end{figure}

We discuss the graph corresponding to the fermion defined on the four-dimensional hyperball.
For simplicity, we first study a graph corresponding to lattice fermion on the discretized one-dimensional ball $B^{1}$ (or line segment).
The digraph corresponding to the discretized $B^{1}$ is given as Fig.~\ref{graph:B1}.
Note that the hopping between two ends in position space is prohibited.
The Dirac matrix corresponding to this graph is given by
\begin{equation}
	\calD_{B^{1}} = Q_{N} \ot \gam{1}
\end{equation}
where $(Q_{N})_{ab} \equiv \left( \sum_{i=1}^{N-1} \delta_{ai} \delta_{i+1\,b} - \sum_{l=2}^{N} \delta_{al}\delta_{l-1\,b} \right)/2$.
This matrix $Q_{N}$ is written as 
\begin{equation}
\label{eq:Qmat}
	Q_{N} = \frac{1}{2}
	\begin{pmatrix}
		0	&1		&0 	&		&0	&0		&0\\
		-1	&0 		&1	&\cdots	&0	&0		&0\\
		0	&-1		&0	&		&0	&0		&0\\
			&\vdots	&	&\ddots	&	&\vdots	&  \\
		0	&0		&0	&		&0	&1		&0\\
		0	&0		&0	&\cdots	&-1	&0		&1\\
		0	&0		&0	&		&0	&-1		&0
	\end{pmatrix}
\end{equation}
The prohibition of hopping in the two ends is represented by the two elements $(1,N)$ and $(N,1)$ which are zero in this case. 

\begin{figure}[htbp]
\centering
	\includegraphics[clip,keepaspectratio, scale=.8]{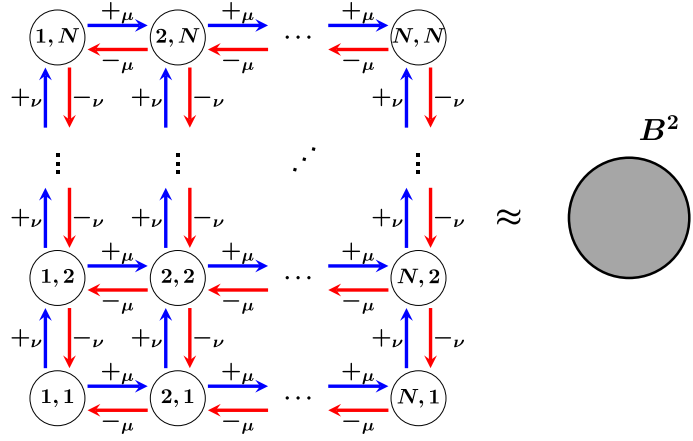}
\caption{This digraph corresponds to fermions on the discretized $B^{2}$. As with the one-dimensional case, the weight of blue edge is $+\gam{\mu}/2$ and the one of the red edge is $-\gam{\mu}/2$.
}
\label{graph:B2}
\end{figure}

For the discretized two-dimensional ball $B^{2}$ (or disk), the digraph is given as Fig.~\ref{graph:B2}.
The Dirac matrix in two dimensions is given by
\begin{equation}
	\calD_{B^{2}} = \ide \ot Q_{N} \ot \gam{1} + Q_{N} \ot \ide \ot \gam{2}.
\end{equation}
As seen from Figs.~\ref{graph:B1} and \ref{graph:B2}, the translational symmetry is partially broken since there is a boundary.

Through a parallel discussion, we find the shape of the digraph corresponding to fermions on the discretized three dimensional ball $B^{3}$ become a box, where the sites in this space is expressed by $(n_{1}, n_{2}, n_{3})$ whith $1 \leq n_{i} \leq N$.
The Dirac matrix on the discretized $B^{3}$ is  given as
\begin{equation}
	\calD_{B^{3}}
	= \ide \ot \ide \ot Q_{N} \ot \gam{1} 
		+ \ide \ot Q_{N} \ot \ide \ot \gam{2} 
		+ Q_{N} \ot \ide \ot \ide \ot \gam{3}.
\end{equation}
In a similar manner we can construct the digraph and the corresponding Dirac matrix on the discretized four dimensional hyperball $B^{4}$.
In the next subsection, we will investigate the lattice action on the discretized $B^{4}$.


\subsection{Lattice action on four-dimensional hyperball
}
The free naive fermion action on four-dimensional discretized hyperball $B^{4}$ is
\begin{equation}
	S_{B^{4}} = \sum_{n} \sum_{\mu} \barpsi_{n} \gam{\mu} D_{\mu} \psi_{n}\,,
\end{equation}
where the lattice sites are $n=(n_{1},n_{2},n_{3},n_{4})$ for $1 \leq n_{i} \leq N$ and Dirichlet boundary conditions are imposed on each direction. This action is also expressed as $S_{B^{4}} = \barpsi \calD_{B^{4}} \psi$. The Dirac matrix $\calD_{B^{4}}$ is
\begin{equation}
\begin{split}
	\calD_{B^{4}}
	&= \ide 	\ot \ide 	\ot \ide 	\ot Q_{N} 	\ot \gam{1} \\
		&+ \ide 	\ot \ide 	\ot Q_{N}	\ot \ide	\ot \gam{2} \\
		&+ \ide	\ot Q_{N}	\ot \ide	\ot \ide	\ot \gam{3} \\
		&+ Q_{N}	\ot \ide	\ot \ide	\ot \ide	\ot \gam{4}.
\end{split}
\end{equation}


\subsection{
Number of species on four-dimensional hyperball
}
Firstly, we will discuss diagonalization of the matrix $Q_{N}$ so as to derive the number of species on $B^{4}$. This matrix can be diagonalized by use of a unitary matrix $Y$ defined as $(Y)_{ab} = \alpha i^{b}\sin \left( \frac{ab\pi}{N+1} \right)$, where $\alpha$ is a normalization coefficient \cite{Noschese2013,Gover1994}. By using the unitary matrix, the diagonalized matrix is given as
\begin{equation}
	Q_{N} Y = i \diag{\cos \left( \frac{\pi}{N+1} \right),\, \cos \left( \frac{2\pi}{N+1} \right),\, \cdots,\, \cos \left( \frac{N\pi}{N+1} \right) } \equiv \Lambda_{Q_{N}} X\,.
\end{equation}
Note that there is a single zero eigenvalue in $\Lambda_{Q_{N}}$ at most when the number of sites in each direction $N$ is odd number. 
The Dirac matrix $\calD_{B^{4}}$ is diagonalized as
\begin{equation}
\begin{split}
	\calV^{\dagger}\, \calD_{B^{4}}\, \calV 
	&= \ide 	\ot \ide 	\ot \ide 	\ot \Lambda_{Q_{N}} 	\ot \gam{1} \\
	&+ \ide 	\ot \ide 	\ot \Lambda_{Q_{N}}	\ot \ide	\ot \gam{2} \\
	&+ \ide	\ot \Lambda_{Q_{N}}	\ot \ide	\ot \ide	\ot \gam{3} \\
	&+ \Lambda_{Q_{N}}	\ot \ide	\ot \ide	\ot \ide	\ot \gam{4}
\end{split} 
\end{equation}
where $\calV$ is a unitary matrix defined as $\calV \equiv \bigot_{\mu=1}^{4} Y \ot \idef$.
To study eigenvalues in the Dirac matrix $\calD_{B^{4}}$, we introduce eigenvector $\ket{\lambda}$ satisfying $Q_{N} \ket{\lambda} = i\cos \left( \frac{\lambda \pi}{N+1} \right) \ket{\lambda}$ for $\lambda = 1,2,\cdots,N$. The diagonalized Dirac matrix is then written
\begin{equation}
	\calV^{\dagger}\, \calD_{B^{4}}\, \calV 
	= \sum_{\lambda_{1},\lambda_{2},\lambda_{3},\lambda_{4}}
	\left[
		i\sum_{\mu=1}^{4} \cos \left( \frac{\lambda_{\mu}\pi}{N+1} \right) \gam{\mu} 
	\right] 
	\ket{\lambda_{1},\lambda_{2},\lambda_{3},\lambda_{4}} \bra{\lambda_{1},\lambda_{2},\lambda_{3},\lambda_{4}}
\end{equation}
where $\ket{\lambda_{1},\lambda_{2},\lambda_{3},\lambda_{4}} = \ket{\lambda_{1}} \ot \ket{\lambda_{2}} \ot \ket{\lambda_{3}} \ot \ket{\lambda_{4}}$.
For the diagonal matrix $\calV^{\dagger}\, \calD_{B^{4}}\, \calV$ to have zero eigenvalues, the following equation must be satisfied,
\begin{equation}
	\sum_{\mu}\cos \left( \frac{\lambda_{\mu}\pi}{N+1} \right) \gam{\mu} = 0\,.
\end{equation}
Furthermore, since $\gamma$ matrices is linearly independent, this equation is rewritten as
\begin{equation}
\label{eq:zero_condition}
	\cos \left( \frac{\lambda_{\mu}\pi}{N+1} \right) = 0.
\end{equation}
The solution of Eq.~\ref{eq:zero_condition} is $\lambda_{\mu} = \frac{N+1}{2}$. 
If the number of sites in the each direction is even, there is no zero eigenvalues in the Dirac matrix. 
As a result, there are no solutions $\lambda_{\mu}$ both satisfying Eq.~\ref{eq:zero_condition} and belonging to the set of integer, $\{1,2,\cdots,N\}$.
On the other hand, when $N$ is the odd number, there is a single zero eigenvalue.
Therefore, there is one physical pole on the bulk of four-dimensional hyperball when the number of sites in each direction is the odd number.
If we take a thermodynamical limit for even $N$, one of the non-zero eigenvalue approaches to zero.
This result is true in any dimensional lattice since $\gamma$ matrices in any dimension is linearly independent.
Thus, lattice fermions on the finite-volume lattice of $d$-dimensional hyperball $B^{d}$ have one physical pole on the bulk.

As is well-known, the lattice fermion defined on a lattice with boundaries can have edge modes on the boundaries. Therefore, the edge mode works to cancel the gauge anomaly at the boundary when we introduce gauge fields or link variables.
The existence of a single Dirac mode on the bulk in the present lattice fermion with boundaries is reasonable as with the case of the domain-wall fermion.

The results we have obtained in this section are again not so novel.
However, we have shown that we can easily find the number of zero eigenvalues or the number of species by obtaining the spectral digraph corresponding to the lattice fermions and taking the procedure of discrete Fourier transformation.


\section{Summary and Discussion}
\label{sec:SD}
In this note, we have introduced a new insight into lattice fermions based on spectral graph theory (SGT).
By use of SGT and discrete Fourier transformation (DFT), we can figure out the number of zero eigenvalues of lattice Dirac operators without considering the momentum space.
Then, we can understand the number of doublers on non-torus lattices with arbitrary topologies.  
The most important fact in this analysis is that a lattice fermion on a certain lattice corresponds to one spectral graph. Then, Dirac operators of lattice fermions can be treated as matrices corresponding to graphs.
Once we have a certain lattice and a Dirac operator defined on the lattice, we can derive the corresponding matrix and find the number of zero eigenvalues by use of the discrete Fourier transformation (DFT), leading to understanding of the number of doublers.
We have applied this procedure to the known fermion formulations including Naive fermion, Wilson fermion and Domain-wall fermion, and discuss the number of doublers in terms of SGT.
As the first non-trivial application, we have studied lattice fermions on the discretized hyperball and discussed the number of fermion species on the bulk. 

We now discuss further application of SGT to lattice fermions;

\begin{enumerate}

\item
Application to discretized sphere:

It is possible to apply SGT to lattice fermions on discretized spheres via the method in Sec. \ref{sec:lf_on_Bd}.
The $d$ dimensional sphere can be realized by eliminating the bulk of $d+1$ dimensional hyperball $B^{d+1}$. 
By using SGT, the Dirac matrix in this situation is simply obtained since what we have to do is to eliminate the matrix elements corresponding to the bulk.
We will discuss this case in the future work soon.

\item
A new theorem on the lattice with arbitrary Euler characteristics:

The method based on SGT can be applied to the lattices with arbitrary Euler characteristics since we simply construct the graphs corresponding discretized manifolds with arbitrary topology.
As a result, we may be able to conjecture a new theorem which tells us the number of species on more generic lattices by studying lattice fermions based on SGT.
One of keys for such a theorem is the topology of the graph itself.
We will also discuss this point in the future work soon.

\item
Novel lattice fermions:

We can propose novel lattice fermions by translating a matrix with desirable properties (minimal zero modes, hermiticity or chirality) to a spectral graph, which corresponds to the lattice fermion.
A fermion obtained by this procedure may correspond to a lattice fermion defined on non-torus lattice.

\end{enumerate}

\newpage


\begin{acknowledgements}
T. M. is grateful to K.~Ohta, S.~Matsuura and S.~Kamata for the fruitful discussion on the relation between  graph theory and lattice field theory.
This work of T. M. is supported by the Japan Society for the Promotion of Science (JSPS) 
Grant-in-Aid for Scientific Research (KAKENHI) Grant Numbers 19K03817 and 18H01217.
\end{acknowledgements}


%
	%


\bibliographystyle{utphys}
\bibliography{./QFT,./refs}

\providecommand{\href}[2]{#2}\begingroup\raggedright\begin{thebibliography}{10}

\bibitem{Wilson:1974sk}
K.~G. Wilson, ``{Confinement of Quarks},''
\href{http://dx.doi.org/10.1103/PhysRevD.10.2445}{{\em Phys. Rev.} {\bfseries
  D10} (1974) 2445--2459}.

\bibitem{Creutz:1980zw}
M.~Creutz, ``{Monte Carlo Study of Quantized SU(2) Gauge Theory},''
\href{http://dx.doi.org/10.1103/PhysRevD.21.2308}{{\em Phys. Rev.} {\bfseries
  D21} (1980) 2308--2315}.

\bibitem{Karsten:1980wd}
L.~H. Karsten and J.~Smit, ``{Lattice Fermions: Species Doubling, Chiral
  Invariance, and the Triangle Anomaly},''
  \href{http://dx.doi.org/10.1016/0550-3213(81)90549-6}{{\em Nucl. Phys.}
  {\bfseries B183} (1981) 103}.
[,495(1980)].

\bibitem{Nielsen:1980rz}
H.~B. Nielsen and M.~Ninomiya, ``{Absence of Neutrinos on a Lattice. 1. Proof
  by Homotopy Theory},''
\href{http://dx.doi.org/10.1016/0550-3213(81)90361-8}{{\em Nucl. Phys.}
  {\bfseries B185} (1981) 20}.

\bibitem{Nielsen:1981xu}
H.~B. Nielsen and M.~Ninomiya, ``{Absence of Neutrinos on a Lattice. 2.
  Intuitive Topological Proof},''
\href{http://dx.doi.org/10.1016/0550-3213(81)90524-1}{{\em Nucl. Phys.}
  {\bfseries B193} (1981) 173--194}.

\bibitem{Nielsen:1981hk}
H.~B. Nielsen and M.~Ninomiya, ``{No Go Theorem for Regularizing Chiral
  Fermions},''
\href{http://dx.doi.org/10.1016/0370-2693(81)91026-1}{{\em Phys. Lett.}
  {\bfseries 105B} (1981) 219--223}.

\bibitem{Wilson:1975id}
K.~G. Wilson, \href{http://dx.doi.org/10.1007/978-1-4613-4208-3_6}{``{Quarks
  and Strings on a Lattice},''} in {\em {New Phenomena in Subnuclear Physics:
  Proceedings, International School of Subnuclear Physics, Erice, Sicily, Jul
  11-Aug 1 1975. Part A}}.
\newblock
1975.
\newblock

\bibitem{Kaplan:1992bt}
D.~B. Kaplan, ``{A Method for simulating chiral fermions on the lattice},''
  \href{http://dx.doi.org/10.1016/0370-2693(92)91112-M}{{\em Phys. Lett.}
  {\bfseries B288} (1992) 342--347},
\href{http://arxiv.org/abs/hep-lat/9206013}{{\ttfamily arXiv:hep-lat/9206013
  [hep-lat]}}.

\bibitem{Shamir:1993zy}
Y.~Shamir, ``{Chiral fermions from lattice boundaries},''
  \href{http://dx.doi.org/10.1016/0550-3213(93)90162-I}{{\em Nucl. Phys.}
  {\bfseries B406} (1993) 90--106},
\href{http://arxiv.org/abs/hep-lat/9303005}{{\ttfamily arXiv:hep-lat/9303005
  [hep-lat]}}.

\bibitem{Furman:1994ky}
V.~Furman and Y.~Shamir, ``{Axial symmetries in lattice QCD with Kaplan
  fermions},'' \href{http://dx.doi.org/10.1016/0550-3213(95)00031-M}{{\em Nucl.
  Phys. B} {\bfseries 439} (1995) 54--78},
  \href{http://arxiv.org/abs/hep-lat/9405004}{{\ttfamily
  arXiv:hep-lat/9405004}}.

\bibitem{Neuberger:1998wv}
H.~Neuberger, ``{More about exactly massless quarks on the lattice},''
  \href{http://dx.doi.org/10.1016/S0370-2693(98)00355-4}{{\em Phys. Lett.}
  {\bfseries B427} (1998) 353--355},
\href{http://arxiv.org/abs/hep-lat/9801031}{{\ttfamily arXiv:hep-lat/9801031
  [hep-lat]}}.

\bibitem{Ginsparg:1981bj}
P.~H. Ginsparg and K.~G. Wilson, ``{A Remnant of Chiral Symmetry on the
  Lattice},''
\href{http://dx.doi.org/10.1103/PhysRevD.25.2649}{{\em Phys. Rev.} {\bfseries
  D25} (1982) 2649}.

\bibitem{Kogut:1974ag}
J.~B. Kogut and L.~Susskind, ``{Hamiltonian Formulation of Wilson's Lattice
  Gauge Theories},''
\href{http://dx.doi.org/10.1103/PhysRevD.11.395}{{\em Phys. Rev.} {\bfseries
  D11} (1975) 395--408}.

\bibitem{Susskind:1976jm}
L.~Susskind, ``{Lattice Fermions},''
\href{http://dx.doi.org/10.1103/PhysRevD.16.3031}{{\em Phys. Rev.} {\bfseries
  D16} (1977) 3031--3039}.

\bibitem{Kawamoto:1981hw}
N.~Kawamoto and J.~Smit, ``{Effective Lagrangian and Dynamical Symmetry
  Breaking in Strongly Coupled Lattice QCD},''.

\bibitem{Sharatchandra:1981si}
H.~S. Sharatchandra, H.~J. Thun, and P.~Weisz, ``{Susskind Fermions on a
  Euclidean Lattice},''
\href{http://dx.doi.org/10.1016/0550-3213(81)90200-5}{{\em Nucl. Phys.}
  {\bfseries B192} (1981) 205--236}.

\bibitem{Golterman:1984cy}
M.~F. Golterman and J.~Smit, ``{Selfenergy and Flavor Interpretation of
  Staggered Fermions},''
  \href{http://dx.doi.org/10.1016/0550-3213(84)90424-3}{{\em Nucl. Phys. B}
  {\bfseries 245} (1984) 61--88}.

\bibitem{Golterman:1985dz}
M.~F. Golterman, ``{STAGGERED MESONS},''
  \href{http://dx.doi.org/10.1016/0550-3213(86)90383-4}{{\em Nucl. Phys. B}
  {\bfseries 273} (1986) 663--676}.

\bibitem{Kilcup:1986dg}
G.~Kilcup and S.~R. Sharpe, ``{A Tool Kit for Staggered Fermions},''
  \href{http://dx.doi.org/10.1016/0550-3213(87)90285-9}{{\em Nucl. Phys. B}
  {\bfseries 283} (1987) 493--550}.

\bibitem{Bietenholz:1999km}
W.~Bietenholz and I.~Hip, ``{The Scaling of exact and approximate
  Ginsparg-Wilson fermions},''
  \href{http://dx.doi.org/10.1016/S0550-3213(99)00477-0}{{\em Nucl. Phys. B}
  {\bfseries 570} (2000) 423--451},
  \href{http://arxiv.org/abs/hep-lat/9902019}{{\ttfamily
  arXiv:hep-lat/9902019}}.

\bibitem{Creutz:2010bm}
M.~Creutz, T.~Kimura, and T.~Misumi, ``{Index Theorem and Overlap Formalism
  with Naive and Minimally Doubled Fermions},''
  \href{http://dx.doi.org/10.1007/JHEP12(2010)041}{{\em JHEP} {\bfseries 12}
  (2010) 041},
\href{http://arxiv.org/abs/1011.0761}{{\ttfamily arXiv:1011.0761 [hep-lat]}}.

\bibitem{Durr:2010ch}
S.~Durr and G.~Koutsou, ``{Brillouin improvement for Wilson fermions},''
  \href{http://dx.doi.org/10.1103/PhysRevD.83.114512}{{\em Phys. Rev. D}
  {\bfseries 83} (2011) 114512},
  \href{http://arxiv.org/abs/1012.3615}{{\ttfamily arXiv:1012.3615 [hep-lat]}}.

\bibitem{Durr:2012dw}
S.~Durr, G.~Koutsou, and T.~Lippert, ``{Meson and Baryon dispersion relations
  with Brillouin fermions},''
  \href{http://dx.doi.org/10.1103/PhysRevD.86.114514}{{\em Phys. Rev. D}
  {\bfseries 86} (2012) 114514},
  \href{http://arxiv.org/abs/1208.6270}{{\ttfamily arXiv:1208.6270 [hep-lat]}}.

\bibitem{Misumi:2012eh}
T.~Misumi, ``{New fermion discretizations and their applications},''
  \href{http://dx.doi.org/10.22323/1.164.0005}{{\em PoS} {\bfseries
  LATTICE2012} (2012) 005},
\href{http://arxiv.org/abs/1211.6999}{{\ttfamily arXiv:1211.6999 [hep-lat]}}.

\bibitem{Cho:2013yha}
Y.-G. Cho, S.~Hashimoto, J.-I. Noaki, A.~Juttner, and M.~Marinkovic,
  ``{$O(a^2)$-improved actions for heavy quarks and scaling studies on quenched
  lattices},'' \href{http://dx.doi.org/10.22323/1.187.0255}{{\em PoS}
  {\bfseries LATTICE2013} (2014) 255},
  \href{http://arxiv.org/abs/1312.4630}{{\ttfamily arXiv:1312.4630 [hep-lat]}}.

\bibitem{Cho:2015ffa}
Y.-G. Cho, S.~Hashimoto, A.~Juttner, T.~Kaneko, M.~Marinkovic, J.-I. Noaki, and
  J.~T. Tsang, ``{Improved lattice fermion action for heavy quarks},''
  \href{http://dx.doi.org/10.1007/JHEP05(2015)072}{{\em JHEP} {\bfseries 05}
  (2015) 072}, \href{http://arxiv.org/abs/1504.01630}{{\ttfamily
  arXiv:1504.01630 [hep-lat]}}.

\bibitem{Durr:2017wfi}
S.~Durr and G.~Koutsou, ``{On the suitability of the Brillouin action as a
  kernel to the overlap procedure},''
  \href{http://arxiv.org/abs/1701.00726}{{\ttfamily arXiv:1701.00726
  [hep-lat]}}.

\bibitem{Adams:2009eb}
D.~H. Adams, ``Theoretical foundation for the index theorem on the lattice with
  staggered fermions,''
  \href{http://dx.doi.org/10.1103/PhysRevLett.104.141602}{{\em Phys.Rev.Lett.}
  {\bfseries 104} (2010) 141602},
  \href{http://arxiv.org/abs/0912.2850}{{\ttfamily arXiv:0912.2850 [hep-lat]}}.

\bibitem{Adams:2010gx}
D.~H. Adams, ``Pairs of chiral quarks on the lattice from staggered fermions,''
  \href{http://dx.doi.org/10.1016/j.physletb.2011.04.034}{{\em Phys.Lett.B}
  {\bfseries 699} (2011) 394--397},
  \href{http://arxiv.org/abs/1008.2833}{{\ttfamily arXiv:1008.2833 [hep-lat]}}.

\bibitem{Hoelbling:2010jw}
C.~Hoelbling, ``Single flavor staggered fermions,''
  \href{http://dx.doi.org/10.1016/j.physletb.2010.12.062}{{\em Phys.Lett.B}
  {\bfseries 696} (2011) 422--425},
  \href{http://arxiv.org/abs/1009.5362}{{\ttfamily arXiv:1009.5362 [hep-lat]}}.

\bibitem{deForcrand:2011ak}
P.~de~Forcrand, A.~Kurkela, and M.~Panero, ``{Numerical properties of staggered
  overlap fermions},'' {\em PoS} {\bfseries LATTICE2010} (2010) 080,
  \href{http://arxiv.org/abs/1102.1000}{{\ttfamily arXiv:1102.1000 [hep-lat]}}.

\bibitem{Creutz:2011cd}
M.~Creutz, T.~Kimura, and T.~Misumi, ``{Aoki Phases in the Lattice Gross-Neveu
  Model with Flavored Mass terms},''
  \href{http://dx.doi.org/10.1103/PhysRevD.83.094506}{{\em Phys. Rev.}
  {\bfseries D83} (2011) 094506},
\href{http://arxiv.org/abs/1101.4239}{{\ttfamily arXiv:1101.4239 [hep-lat]}}.

\bibitem{Misumi:2011su}
T.~Misumi, M.~Creutz, T.~Kimura, T.~Z. Nakano, and A.~Ohnishi, ``{Aoki Phases
  in Staggered-Wilson Fermions},''
  \href{http://dx.doi.org/10.22323/1.139.0108}{{\em PoS} {\bfseries
  LATTICE2011} (2011) 108}, \href{http://arxiv.org/abs/1110.1231}{{\ttfamily
  arXiv:1110.1231 [hep-lat]}}.

\bibitem{Follana:2011kh}
E.~Follana, V.~Azcoiti, G.~Di~Carlo, and A.~Vaquero, ``{Spectral Flow and Index
  Theorem for Staggered Fermions},''
  \href{http://dx.doi.org/10.22323/1.139.0100}{{\em PoS} {\bfseries
  LATTICE2011} (2011) 100}, \href{http://arxiv.org/abs/1111.3502}{{\ttfamily
  arXiv:1111.3502 [hep-lat]}}.

\bibitem{deForcrand:2012bm}
P.~de~Forcrand, A.~Kurkela, and M.~Panero, ``Numerical properties of staggered
  quarks with a taste-dependent mass term,''
  \href{http://dx.doi.org/10.1007/JHEP04(2012)142}{{\em JHEP} {\bfseries 04}
  (2012) 142}, \href{http://arxiv.org/abs/1202.1867}{{\ttfamily arXiv:1202.1867
  [hep-lat]}}.

\bibitem{Misumi:2012sp}
T.~Misumi, T.~Z. Nakano, T.~Kimura, and A.~Ohnishi, ``Strong-coupling analysis
  of parity phase structure in staggered-wilson fermions,''
  \href{http://dx.doi.org/10.1103/PhysRevD.86.034501}{{\em Phys.Rev.D}
  {\bfseries 86} (2012) 034501},
  \href{http://arxiv.org/abs/1205.6545}{{\ttfamily arXiv:1205.6545 [hep-lat]}}.

\bibitem{Durr:2013gp}
S.~Durr, ``{Taste-split staggered actions: eigenvalues, chiralities and
  Symanzik improvement},''
  \href{http://dx.doi.org/10.1103/PhysRevD.87.114501}{{\em Phys. Rev. D}
  {\bfseries 87} no.~11, (2013) 114501},
  \href{http://arxiv.org/abs/1302.0773}{{\ttfamily arXiv:1302.0773 [hep-lat]}}.

\bibitem{Hoelbling:2016qfv}
C.~Hoelbling and C.~Zielinski, ``{Spectral properties and chiral symmetry
  violations of (staggered) domain wall fermions in the Schwinger model},''
  \href{http://dx.doi.org/10.1103/PhysRevD.94.014501}{{\em Phys. Rev. D}
  {\bfseries 94} no.~1, (2016) 014501},
  \href{http://arxiv.org/abs/1602.08432}{{\ttfamily arXiv:1602.08432
  [hep-lat]}}.

\bibitem{Zielinski:2017pko}
C.~Zielinski, {\em {Theoretical and Computational Aspects of New Lattice
  Fermion Formulations}}.
\newblock PhD thesis, Nanyang Technol. U., 2016.
\newblock \href{http://arxiv.org/abs/1703.06364}{{\ttfamily arXiv:1703.06364
  [hep-lat]}}.

\bibitem{Karsten:1981gd}
L.~H. Karsten, ``{Lattice Fermions in Euclidean Space-time},''
  \href{http://dx.doi.org/10.1016/0370-2693(81)90133-7}{{\em Phys. Lett. B}
  {\bfseries 104} (1981) 315--319}.

\bibitem{Wilczek:1987kw}
F.~Wilczek, ``{ON LATTICE FERMIONS},''
  \href{http://dx.doi.org/10.1103/PhysRevLett.59.2397}{{\em Phys. Rev. Lett.}
  {\bfseries 59} (1987) 2397}.

\bibitem{Creutz:2007af}
M.~Creutz, ``{Four-dimensional graphene and chiral fermions},''
  \href{http://dx.doi.org/10.1088/1126-6708/2008/04/017}{{\em JHEP} {\bfseries
  04} (2008) 017}, \href{http://arxiv.org/abs/0712.1201}{{\ttfamily
  arXiv:0712.1201 [hep-lat]}}.

\bibitem{Borici:2007kz}
A.~Borici, ``{Creutz fermions on an orthogonal lattice},''
  \href{http://dx.doi.org/10.1103/PhysRevD.78.074504}{{\em Phys. Rev. D}
  {\bfseries 78} (2008) 074504},
  \href{http://arxiv.org/abs/0712.4401}{{\ttfamily arXiv:0712.4401 [hep-lat]}}.

\bibitem{Bedaque:2008xs}
P.~F. Bedaque, M.~I. Buchoff, B.~C. Tiburzi, and A.~Walker-Loud, ``{Broken
  Symmetries from Minimally Doubled Fermions},''
  \href{http://dx.doi.org/10.1016/j.physletb.2008.03.034}{{\em Phys. Lett. B}
  {\bfseries 662} (2008) 449--455},
  \href{http://arxiv.org/abs/0801.3361}{{\ttfamily arXiv:0801.3361 [hep-lat]}}.

\bibitem{Bedaque:2008jm}
P.~F. Bedaque, M.~I. Buchoff, B.~C. Tiburzi, and A.~Walker-Loud, ``{Search for
  Fermion Actions on Hyperdiamond Lattices},''
  \href{http://dx.doi.org/10.1103/PhysRevD.78.017502}{{\em Phys. Rev. D}
  {\bfseries 78} (2008) 017502},
  \href{http://arxiv.org/abs/0804.1145}{{\ttfamily arXiv:0804.1145 [hep-lat]}}.

\bibitem{Capitani:2009yn}
S.~Capitani, J.~Weber, and H.~Wittig, ``{Minimally doubled fermions at one
  loop},'' \href{http://dx.doi.org/10.1016/j.physletb.2009.09.050}{{\em Phys.
  Lett. B} {\bfseries 681} (2009) 105--112},
  \href{http://arxiv.org/abs/0907.2825}{{\ttfamily arXiv:0907.2825 [hep-lat]}}.

\bibitem{Kimura:2009qe}
T.~Kimura and T.~Misumi, ``{Characters of Lattice Fermions Based on the
  Hyperdiamond Lattice},'' \href{http://dx.doi.org/10.1143/PTP.124.415}{{\em
  Prog. Theor. Phys.} {\bfseries 124} (2010) 415--432},
  \href{http://arxiv.org/abs/0907.1371}{{\ttfamily arXiv:0907.1371 [hep-lat]}}.

\bibitem{Kimura:2009di}
T.~Kimura and T.~Misumi, ``{Lattice Fermions Based on Higher-Dimensional
  Hyperdiamond Lattices},'' \href{http://dx.doi.org/10.1143/PTP.123.63}{{\em
  Prog. Theor. Phys.} {\bfseries 123} (2010) 63--78},
  \href{http://arxiv.org/abs/0907.3774}{{\ttfamily arXiv:0907.3774 [hep-lat]}}.

\bibitem{Creutz:2010cz}
M.~Creutz and T.~Misumi, ``{Classification of Minimally Doubled Fermions},''
  \href{http://dx.doi.org/10.1103/PhysRevD.82.074502}{{\em Phys. Rev. D}
  {\bfseries 82} (2010) 074502},
  \href{http://arxiv.org/abs/1007.3328}{{\ttfamily arXiv:1007.3328 [hep-lat]}}.

\bibitem{Capitani:2010nn}
S.~Capitani, M.~Creutz, J.~Weber, and H.~Wittig, ``{Renormalization of
  minimally doubled fermions},''
  \href{http://dx.doi.org/10.1007/JHEP09(2010)027}{{\em JHEP} {\bfseries 09}
  (2010) 027}, \href{http://arxiv.org/abs/1006.2009}{{\ttfamily arXiv:1006.2009
  [hep-lat]}}.

\bibitem{Tiburzi:2010bm}
B.~C. Tiburzi, ``{Chiral Lattice Fermions, Minimal Doubling, and the Axial
  Anomaly},'' \href{http://dx.doi.org/10.1103/PhysRevD.82.034511}{{\em Phys.
  Rev. D} {\bfseries 82} (2010) 034511},
  \href{http://arxiv.org/abs/1006.0172}{{\ttfamily arXiv:1006.0172 [hep-lat]}}.

\bibitem{Kamata:2011jn}
S.~Kamata and H.~Tanaka, ``{Minimal Doubling Fermion and Hermiticity},''
  \href{http://dx.doi.org/10.1093/ptep/pts093}{{\em PTEP} {\bfseries 2013}
  (2013) 023B05}, \href{http://arxiv.org/abs/1111.4536}{{\ttfamily
  arXiv:1111.4536 [hep-lat]}}.

\bibitem{Misumi:2012uu}
T.~Misumi, ``{Phase structure for lattice fermions with flavored chemical
  potential terms},'' \href{http://dx.doi.org/10.1007/JHEP08(2012)068}{{\em
  JHEP} {\bfseries 08} (2012) 068},
  \href{http://arxiv.org/abs/1206.0969}{{\ttfamily arXiv:1206.0969 [hep-lat]}}.

\bibitem{Misumi:2012ky}
T.~Misumi, T.~Kimura, and A.~Ohnishi, ``{QCD phase diagram with 2-flavor
  lattice fermion formulations},''
  \href{http://dx.doi.org/10.1103/PhysRevD.86.094505}{{\em Phys. Rev. D}
  {\bfseries 86} (2012) 094505},
  \href{http://arxiv.org/abs/1206.1977}{{\ttfamily arXiv:1206.1977 [hep-lat]}}.

\bibitem{Capitani:2013zta}
S.~Capitani, ``{Reducing the number of counterterms with new minimally doubled
  actions},'' \href{http://dx.doi.org/10.1103/PhysRevD.89.014501}{{\em Phys.
  Rev. D} {\bfseries 89} no.~1, (2014) 014501},
  \href{http://arxiv.org/abs/1307.7497}{{\ttfamily arXiv:1307.7497 [hep-lat]}}.

\bibitem{Capitani:2013iha}
S.~Capitani, ``{New chiral lattice actions of the Borici-Creutz type},''
  \href{http://dx.doi.org/10.1103/PhysRevD.89.074508}{{\em Phys. Rev. D}
  {\bfseries 89} no.~7, (2014) 074508},
  \href{http://arxiv.org/abs/1311.5664}{{\ttfamily arXiv:1311.5664 [hep-lat]}}.

\bibitem{Misumi:2013maa}
T.~Misumi, ``{Fermion Actions extracted from Lattice Super Yang-Mills
  Theories},'' \href{http://dx.doi.org/10.1007/JHEP12(2013)063}{{\em JHEP}
  {\bfseries 12} (2013) 063}, \href{http://arxiv.org/abs/1311.4365}{{\ttfamily
  arXiv:1311.4365 [hep-lat]}}.

\bibitem{Weber:2013tfa}
J.~H. Weber, S.~Capitani, and H.~Wittig, ``{Numerical studies of Minimally
  Doubled Fermions},'' \href{http://dx.doi.org/10.22323/1.187.0122}{{\em PoS}
  {\bfseries LATTICE2013} (2014) 122},
  \href{http://arxiv.org/abs/1312.0488}{{\ttfamily arXiv:1312.0488 [hep-lat]}}.

\bibitem{Weber:2017eds}
J.~H. Weber, {\em {Properties of minimally doubled fermions}}.
\newblock PhD thesis, Mainz U., 2015.
\newblock \href{http://arxiv.org/abs/1706.07104}{{\ttfamily arXiv:1706.07104
  [hep-lat]}}.

\bibitem{Durr:2020yqa}
S.~Durr and J.~H. Weber, ``{Dispersion relation and spectral range of
  Karsten-Wilczek and Borici-Creutz fermions},''
  \href{http://arxiv.org/abs/2003.10803}{{\ttfamily arXiv:2003.10803
  [hep-lat]}}.

\bibitem{Kimura:2011ik}
T.~Kimura, S.~Komatsu, T.~Misumi, T.~Noumi, S.~Torii, and S.~Aoki,
  ``{Revisiting symmetries of lattice fermions via spin-flavor
  representation},'' \href{http://dx.doi.org/10.1007/JHEP01(2012)048}{{\em
  JHEP} {\bfseries 01} (2012) 048},
\href{http://arxiv.org/abs/1111.0402}{{\ttfamily arXiv:1111.0402 [hep-lat]}}.

\bibitem{Chowdhury:2013ux}
A.~Chowdhury, A.~Harindranath, J.~Maiti, and S.~Mondal, ``Many avatars of the
  wilson fermion: A perturbative analysis,''
  \href{http://dx.doi.org/10.1007/JHEP02(2013)037}{{\em JHEP} {\bfseries 02}
  (2013) 037}, \href{http://arxiv.org/abs/1301.0675}{{\ttfamily arXiv:1301.0675
  [hep-lat]}}.

\bibitem{Misumi:2020eyx}
T.~Misumi and J.~Yumoto, ``{Varieties and properties of central-branch Wilson
  fermions},'' \href{http://dx.doi.org/10.1103/PhysRevD.102.034516}{{\em Phys.
  Rev. D} {\bfseries 102} no.~3, (2020) 034516},
  \href{http://arxiv.org/abs/2005.08857}{{\ttfamily arXiv:2005.08857
  [hep-lat]}}.

\bibitem{west2001introduction}
D.~West, {\em Introduction to Graph Theory}.
\newblock Featured Titles for Graph Theory. Prentice Hall, 2001.
\newblock \url{https://books.google.co.jp/books?id=TuvuAAAAMAAJ}.

\bibitem{bondy1976graph}
J.~Bondy and U.~Murty, {\em Graph Theory with Applications}.
\newblock American Elsevier Publishing Company, 1976.
\newblock \url{https://books.google.co.jp/books?id=4bwrAAAAYAAJ}.

\bibitem{mieghem_2010}
P.~v. Mieghem, \href{http://dx.doi.org/10.1017/CBO9780511921681}{{\em Graph
  Spectra for Complex Networks}}.
\newblock Cambridge University Press, 2010.

\bibitem{Watts1998}
D.~J. Watts and S.~H. Strogatz, ``Collective dynamics of ‘small-world’
  networks,'' \href{http://dx.doi.org/10.1038/30918}{{\em Nature} {\bfseries
  393} no.~6684, (1998) 440--442}. \url{https://doi.org/10.1038/30918}.

\bibitem{Ohta:2020ygi}
K.~Ohta and N.~Sakai, ``{The Volume of the Quiver Vortex Moduli Space},''
  \href{http://dx.doi.org/10.1093/ptep/ptab012}{{\em PTEP} {\bfseries 2021}
  no.~3, (2021) 033B02}, \href{http://arxiv.org/abs/2009.09580}{{\ttfamily
  arXiv:2009.09580 [hep-th]}}.

\bibitem{Ohta:2021xty}
K.~Ohta and S.~Matsuura, ``{Supersymmetric Gauge Theory on the Graph},''
  \href{http://arxiv.org/abs/2111.00676}{{\ttfamily arXiv:2111.00676
  [hep-th]}}.

\bibitem{Catterall:2018lkj}
S.~Catterall, J.~Laiho, and J.~Unmuth-Yockey, ``{Topological fermion
  condensates from anomalies},''
  \href{http://dx.doi.org/10.1007/JHEP10(2018)013}{{\em JHEP} {\bfseries 10}
  (2018) 013}, \href{http://arxiv.org/abs/1806.07845}{{\ttfamily
  arXiv:1806.07845 [hep-lat]}}.

\bibitem{Butt:2021brl}
N.~Butt, S.~Catterall, A.~Pradhan, and G.~C. Toga, ``{Anomalies and symmetric
  mass generation for K\"ahler-Dirac fermions},''
  \href{http://dx.doi.org/10.1103/PhysRevD.104.094504}{{\em Phys. Rev. D}
  {\bfseries 104} no.~9, (2021) 094504},
  \href{http://arxiv.org/abs/2101.01026}{{\ttfamily arXiv:2101.01026
  [hep-th]}}.

\bibitem{Kaveh2005}
A.~Kaveh and H.~Rahami, ``A unified method for eigendecomposition of graph
  products,'' \href{http://dx.doi.org/https://doi.org/10.1002/cnm.753}{{\em
  Communications in Numerical Methods in Engineering} {\bfseries 21} no.~7,
  (2005) 377--388},
  \href{http://arxiv.org/abs/https://onlinelibrary.wiley.com/doi/pdf/10.1002/cnm.753}{{\ttfamily
  https://onlinelibrary.wiley.com/doi/pdf/10.1002/cnm.753}}.
  \url{https://onlinelibrary.wiley.com/doi/abs/10.1002/cnm.753}.

\bibitem{Sabidussi1959}
G.~Sabidussi, ``Graph multiplication,''
  \href{http://dx.doi.org/10.1007/BF01162967}{{\em Mathematische Zeitschrift}
  {\bfseries 72} no.~1, (1959) 446--457}.
  \url{https://doi.org/10.1007/BF01162967}.

\bibitem{Aurenhammer1992}
F.~Aurenhammer, J.~Hagauer, and W.~Imrich, ``Cartesian graph factorization at
  logarithmic cost per edge,'' \href{http://dx.doi.org/10.1007/BF01200428}{{\em
  computational complexity} {\bfseries 2} no.~4, (1992) 331--349}.
  \url{https://doi.org/10.1007/BF01200428}.

\bibitem{Noschese2013}
S.~Noschese, L.~Pasquini, and L.~Reichel, ``Tridiagonal toeplitz matrices:
  properties and novel applications,''
  \href{http://dx.doi.org/https://doi.org/10.1002/nla.1811}{{\em Numerical
  Linear Algebra with Applications} {\bfseries 20} no.~2, (2013) 302--326},
  \href{http://arxiv.org/abs/https://onlinelibrary.wiley.com/doi/pdf/10.1002/nla.1811}{{\ttfamily
  https://onlinelibrary.wiley.com/doi/pdf/10.1002/nla.1811}}.
  \url{https://onlinelibrary.wiley.com/doi/abs/10.1002/nla.1811}.

\bibitem{Gover1994}
M.~Gover, ``The eigenproblem of a tridiagonal 2-toeplitz matrix,''
  \href{http://dx.doi.org/10.1016/0024-3795(94)90481-2}{{\em Linear Algebra and
  its Applications} {\bfseries 197-198} (02, 1994) 63--78}.

\end{thebibliography}\endgroup

\end{document}